\documentclass[preprint]{elsarticle}
\usepackage{graphicx}
\usepackage{bm}
\usepackage{float}
\usepackage{physics}
\usepackage{amssymb}

\begin{document}
	
	\begin{frontmatter}
		\title{Exact Semiclassical Dynamics in a Multidimensional Quartic Potential: Multi-Flavor Instantons and Pre-Factor Dominance at Strong Coupling}
		
		\author[1]{Pervez Hoodbhoy\corref{cor1}}
		\ead{hoodbhoy@mit.edu}
		\author[1]{M. Haashir Ismail}
		\author[1]{M. Mufassir}
		
		\address[1]{The Black Hole, Sector G-11/3, Islamabad, Pakistan}
		\cortext[cor1]{Corresponding author}
		
\begin{abstract}
We present an exact semi-classical analysis of a coupled multi-degree-of-freedom quantum system governed by a symmetric quartic potential with four degenerate minima.  Starting from the Euclidean path integral, we identify distinct paths—longitudinal, transverse, and diagonal—corresponding to distinct instanton configurations that mediate tunneling between vacua. The continuous translational zero mode for each coupled trajectory is extracted rigorously by transforming the fluctuation operator into a comoving rotating frame, explicitly incorporating curvature-induced fictitious forces. By evaluating the exact functional determinants via the Gelfand-Yaglom method and mapping the multi-flavor dilute instanton gas onto a weighted $K_4$ adjacency matrix, we analytically derive the transition amplitudes and ground-state energy splittings. Crucially, our exact continuum treatment reveals that for strong coupling, the usual exponential can be overpowered in certain parameter ranges by the "pre-factor". This quantity, which arises from quadratic level fluctuations in the direction transverse to the free direction, turns out to have an essential singularity that would be hard to spot in standard discretized instanton approximations.
		\end{abstract}
		\begin{keyword}
			multidimensional tunneling \sep instanton theory \sep semiclassical analysis \sep functional determinants \sep rotating frame \sep path integrals \sep zero modes
			
			\vspace{0.5cm} 
			\textbf{PACS:} 03.65.Sq \sep 03.65.Xp \sep 11.15.Kc \sep 02.30.Mv
		\end{keyword}
	\end{frontmatter}
	
	\section{Introduction}
	The double-well potential was the paradigmatic starting point for investigating quantum tunneling between degenerate vacuum states that served to launch seminal developments in instanton calculus during the 1970s \cite{Novikov, Coleman2,tHooft}. Thereafter non-perturbative semiclassical methods became indispensable for uncovering the topological properties of Yang-Mills gauge theories, the spectral structures of quantum systems, and the mechanics of false vacuum decay. These early developments have been reviewed by several authors \cite{Devoto,Shifman, Marino,Teper}.
	In later decades, the instanton concept found application across fields such as discrete statistical formulations of quantum mechanics and lattice studies of false vacuum decay \cite{Creutz,Munster,Kogut,Bender1, Batini}.
	The basic instanton model was then generalized to multidimensional spaces where multiple fields are mutually coupled such as the cosmological Two Higgs Doublet Models (2HDM) \cite{Branchina}.
	
	Given the ubiquity of instanton calculus, it is rather surprising that the present investigation is the first to extend the canonical instanton framework to multidimensional systems wherein two or more fields are mutually coupled. Part of the reason behind the delay may well be due to a purely technical issue: the classical equations of motion are coupled nonlinear differential equations that very rarely admit closed-form solutions. 
	
The bigger difficulty is conceptual: it is far from clear how zero modes can be handled, contributions from multiple classical paths can be summed, and the fluctuation determinants evaluated. The goal of this paper is to bridge an analytical gap between the single DOF, two-minima model to two DOFs and four identical minima. We take a stripped down solvable ``toy model'' that nevertheless preserves critical features common to all higher dimensional models. Even in such a simplified system, tunneling is not a solitary event but a collective topological process. Energy conservation mandates that interacting fields must ``lock'' together to traverse their respective barriers synchronously. For this they must concentrate their combined energy within a highly localized temporal window. This intimately connects with the system's particular path - the zero mode - along which all fields can be moved simultaneously without paying a price in energy.
	
	The zero mode problem is tackled here by transforming the coupled system into a comoving rotating frame. This isolates the longitudinal, ``force-free'' tunneling direction from the transverse degrees of freedom. After exploring the stability of transverse fluctuations in this frame—which accounts for Coriolis and velocity-dependent fictitious forces—we analytically evaluate the functional determinants using the Gelfand-Yaglom theorem \cite{Gelfand}, which becomes the prefactor. 
	In evaluating the Jacobian, we use (see Appendix A) the Faddeev-Popov \cite{FP1,FP2} procedure, which was applied to the single DOF case by Zinn-Justin \cite{Zinn}. While this evaluation could be done more straightforwardly, we used this as a safety precaution to foreclose the possibility of an ambiguity.
	
	A second conceptual hurdle is that of summing over instanton contributions. We recall from the 1-D double well that the amplitude is a product of the zero mode volume (proportional to $T$) and an exponential $e^{-\omega T}$. Hence in the limit $T\rightarrow \infty$ the contribution of a single instanton to $\mathcal{A}$ vanishes identically. This led to the invention of the dilute gas non-overlapping instanton model wherein a coherent sum is performed over an infinite number of time-sequenced instantons, and hence a finite result in the $T\rightarrow \infty$ limit. The multi-dimensional case is similar except that here the number and complexity of contributing paths is vastly greater. To tackle this, we introduce a graph-theoretic summation which maps the competing multi-flavor instantons (edge and diagonal transitions) onto a $K_4$ adjacency matrix. Analytically executing the infinite sum over all interacting topological pathways, we derive exact closed-form expressions for the coherent Rabi-type oscillations and the low-lying energy splittings. Because of the simplicity of our model, solving the 2-d Schr\"{o}dinger equation to high accuracy is straightforward using existing PDE solvers. Comparing energy splittings provides a vital check on the correctness of our formalism.
	
	An unanticipated consequence of our investigation was that the prefactor - which is generally assumed to be subdominant - can develop an essential singularity for strong enough attractive coupling and this can swamp the usual $\exp(-S_0)$ tunneling factor. At this point the semi-classical approach collapses. This happens because the potential manifold no longer supports tunneling solutions.
	
	While the model explored here allows for some interesting theoretical insights, we recognize the limitations of a purely analytical approach. Even for our toy model, we had to make certain parameter choices to get closed form answers. To solve practical problems, one has no choice but to evaluate quantum mechanical path integrals by mapping them on to statistical mechanics models, a step that fundamentally relies on the discretization of Euclidean time on a finite lattice. Other than lattice gauge theory, discretized instanton formulations—such as Ring Polymer Instanton (RPI) models—are widely used in chemical physics to compute energy splittings and chemical reaction rates \cite{Richardson1, Richardson2011, Lawrence2023, Richardson3, Mano, Kast, Kry,Erak, Benderskii}. In Section 9 we shall briefly compare and contrast RPI with the present analytical (continuum) formulation. While the starting point is identical and some basic similarities are evident, the relationship is by no means simple; much remains to be understood. The bottom line: analytical explorations - such as the present one - are much needed.
	
	\section{Coupled instantons model}
\begin{figure}[H]
		\centering
		\hspace*{-0.5cm} 
		\includegraphics[scale=.40]{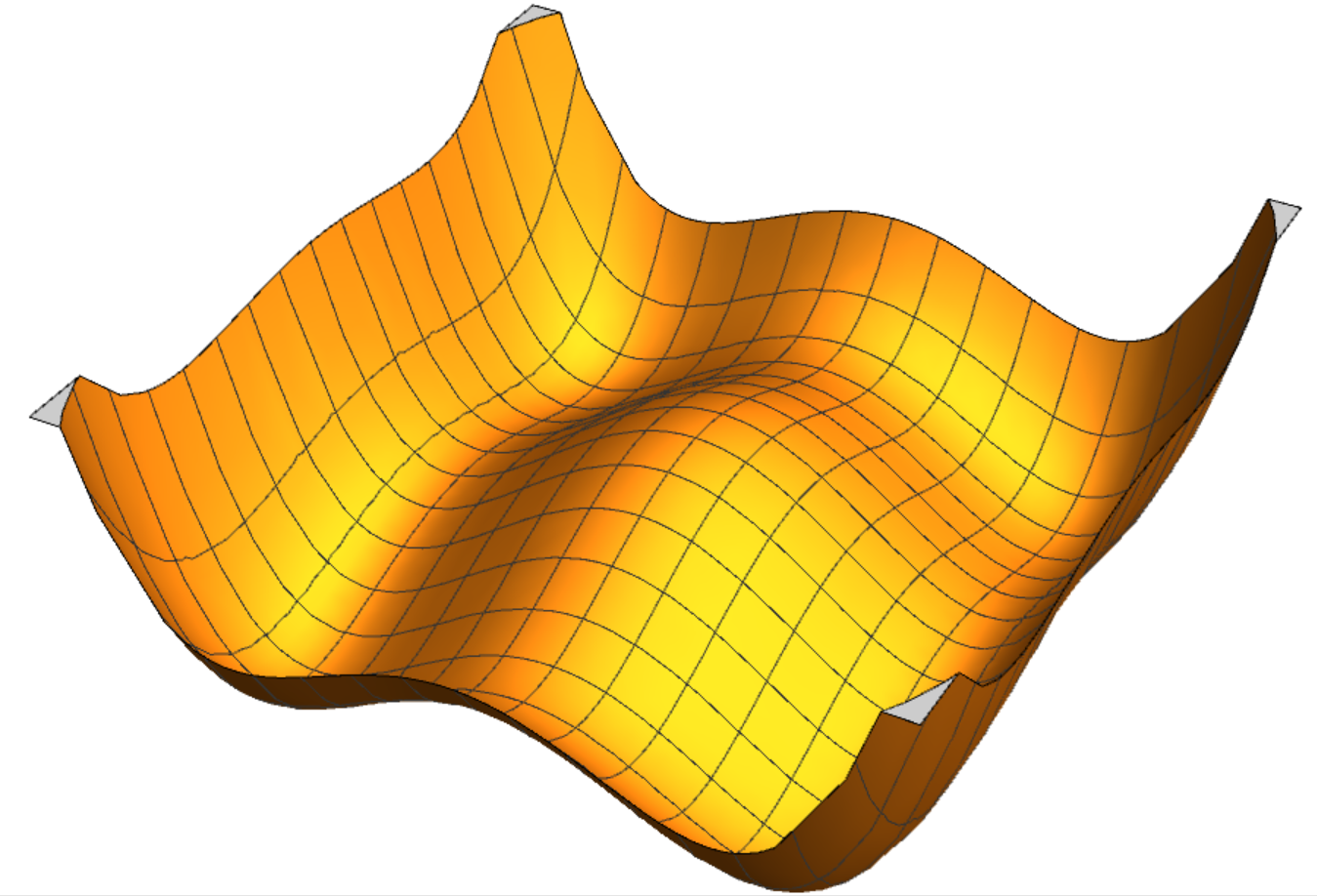}
		\caption{Schematic of the four-well symmetric quartic potential $V(p,q)$ for parameters $b_p=2, b_q=1,$ and $c=1/2$.
			The four degenerate minima at $(\pm 1, \pm 1)$ are separated by saddle points, facilitating multiple distinct tunneling pathways.}\label{potential}
	\end{figure}

	All tunneling calculations ultimately seek to find the Feynman amplitude in imaginary (or Euclidean) time $t$, 
	\begin{eqnarray}		
		\mathcal{A}_{if}=\big\langle f,T/2| e^{-\frac{Ht}{\hbar}} |i,-T/2\big\rangle=
		\mathcal{N}\int[dx(t)]\; e^{-S[x]},
		\label{amp}	
	\end{eqnarray}
where $S[x]$ is the dimensionless  Euclidean action. 
To be kept in mind is the smallness of $\hbar$; this is key to the semi-classical method's success.The system is in its initial state at time $t=-T/2$ and makes its way to the final state at $t=T/2$ by traveling on all possible paths connecting the initial to the final state. 
	To get bound state energies and wavefunctions one eventually takes $T\rightarrow\infty$.
	In the semiclassical approximation, fluctuations away from these paths are allowed up to the quadratic level while the rest can be treated in a perturbative expansion.
	Building on the techniques developed in the foundational works of instanton theory \cite{Novikov, Coleman2, tHooft} (a recent review \cite{Devoto} is particularly useful), in this paper we shall investigate the system defined by the Euclidean Lagrangian, 
	\begin{eqnarray}
		L&=&\frac{1}{2}m_p\left(\frac{dx}{dt}\right)^2 +
		\frac{1}{2}m_q\left(\frac{dy}{dt}\right)^2+V(x,y)
		\nonumber \\ V(x,y)&=&\frac{m_p\omega_p^2}{8x_p^2}(x^2-x_p^2)^2 + \frac{m_q\omega_q^2}{8y_q^2}(y^2-y_q^2)^2 +c_{pq}(x^2-x_p^2)(y^2-y_q^2).
	\end{eqnarray}	
	Standard notation has been used to represent two symmetric double wells with minima at $x_p,y_q$ coupled together with strength $c_{pq}$.
	We make the following re-definitions:
	\begin{eqnarray}		
		p=\frac{x}{x_p},\;\;q=\frac{y}{y_q},\;\;
		a_p=m_px_p^2,\;\;
		b_p=m_p x_p^2\omega_p^2,\;\;a_q=m_qy_q^2,\;\;b_q=m_q y_q^2\omega_q^2.	\label{abc}
	\end{eqnarray}
 The five independent constants $a_p,b_p,a_q,b_q,c$ are set by some underlying physical model.
	The particular form of interaction is, as we show in Appendix B, of interest in at least one physical problem but, apart from the tunneling of a composite system of particles, other optical or condensed matter systems could also have such an effective potential.
	The system is defined by the dimensionless action $S$,
	\begin{eqnarray}
		S&=&\int \mathcal{L} dt,
		\quad
		\mathcal{L}=\frac{1}{2}a_p\dot p^2 +\frac{1}{2}a_q\dot q^2 +V(p,q),\nonumber\\
		V(p,q)&=&\frac{1}{8}b_p(p^2-1)^2 +\frac{1}{8}b_q(q^2-1)^2 +\frac{1}{4}c(p^2-1)(q^2-1).
		\label{Pot}
	\end{eqnarray}
	$V(p,q)$ is shown schematically in Fig.~\ref{potential}.
	Consider now two instantons corresponding to two different physical variables interacting with each other through $V(p,q)$.
	Let us first explore the parts of parameter space relevant to the tunneling problem at hand.
	All masses and spring constants are taken positive definite, i.e. $a_p>0,b_p>0,a_q>0,b_q>0$.
	Further, the minima $x_p, y_q$ must satisfy $x_p^2>0,y_q^2>0.$ 
	With $P=p^2-1$ and $Q=q^2-1$ we may rewrite the potential in Eq.~\ref{Pot},
	\begin{eqnarray}
		V(p,q)=\frac{1}{8}b_pP^2 +\frac{1}{8}b_qQ^2 + \frac{1}{4}cPQ=\frac{1}{8}b_p\left(P+\frac{c}{b_p}Q\right)^2 +\frac{\Delta}{8b_p}Q^2.
	\end{eqnarray}
	If the discriminant $\Delta=b_pb_q-c^2 >0$, then $V\ge 0$. 
	The critical points of $V$ are at $A=(\pm1,\pm1), B=(0,q),$ and, $C=(p,0)$.
	To ascertain the behaviour of $V$ we must examine the Hessian matrix $H(p,q)$ at all such points.
	At $A$, $V(A)=0$ and det$[H(A)]=4\Delta>0$. Hence the four $A$ points are indeed true local minima.
	At points $B$,
	\begin{eqnarray}
		V(0,q)&=&\frac{1}{8}b_p +\frac{1}{8}b_qQ^2 - \frac{1}{4}cQ=\frac{1}{8}b_q\left(Q-\frac{c}{b_q}\right)^2 +\frac{\Delta}{8b_q}.
	\end{eqnarray}
	This will take its minimum value $\Delta/(8b_q)>0$
	when $Q_m=c/b_q$ i.e. when $q_m^2=1+c/b_q$.
	At this point det$[H(0,q_m)]=-(1+c/b_q)\Delta.$ 
	This will be a saddle point if $b_q+c>0$. 
	The conclusion for all points $C$ is similar.
	In summary, to have exactly four degenerate minima and no other local minimum, requires:
	\begin{equation}
		b_pb_q>c^2,\;b_p+c>0,\;b_q+c>0.
		\label{cond}
	\end{equation}
	While $V$ is unbounded above in the $R_2$ plane, a local maximum at the origin, 
	\begin{equation}
		V(0,0) =\frac{1}{8}b_p+\frac{1}{8}b_q +\frac{1}{4}c,
	\end{equation}
	is guaranteed because det$[H(0,0)]\sim (b_p+c)(b_q+c)>0$ as per Eq.~\ref{cond}.
	Let us now consider the interaction between the $p$ and $q$ fields.
	This operates only for times when $p^2\ne 1$ and $q^2\ne 1$ and then rapidly turns itself off.
	For $c=0$ two decoupled instantons mediate between their respective vacuum states at $p=\pm1$
	and $q=\pm1$.
	As per the semi-classical prescription, the action $S[p,q]$ in Eq.~\ref{Pot} must be expanded in a functional
	Taylor series around the classical path up to and including quadratic terms, 
	$S=S_0+S_1+S_2+\cdots$.
	What remains can be treated perturbatively. 
	\begin{align}
		S[p,q] = S[p_c,q_c] + \int dt' \left[ \frac{\delta S}{\delta p}\eta + \frac{\delta S}{\delta q}\xi \right] 
		+ \frac{1}{2} \int dt'dt'' \left[ \frac{\delta^2 S}{\delta p\delta p} \eta\eta + \frac{\delta^2 S}{\delta q\delta q} \xi\xi + 2 \frac{\delta^2 S}{\delta p\delta q} \eta\xi + \cdots \right] \label{Squad}
	\end{align}
	The partial derivatives indicated above are evaluated at $p=p_c, q=q_c$.
	Deviations from the classical path $\eta(t)=p-p_c$ and $\xi(t)=q-q_c$ are integrated over subject to the end point conditions, 
	$\eta(- T/2)= \eta(T/2)=\xi(- T/2)=\xi(T/2)=0.$ 
	Some algebra gives the order by order decomposition of the dimensionless action,
	\begin{eqnarray}
		S_0&=&\frac{1}{2} \int dt \left[
		a_p\dot p_c^2
		+ \frac{1}{4}b_p (p_c^2-1)^2 +a_q \dot q_c^2+\frac{1}{4}b_q(q_c^2-1)^2
		+ \frac{1}{2}c (p_c^2-1)(q_c^2-1)
		\right],
		\nonumber\\
		S_1&=&\int dt
		\left[ a_p\dot p_c\dot\eta 
		+\frac{b_p}{2}p_c(p_c^2-1)\eta+
		\frac{c}{2}p_c(q_c^2-1)
		\eta+a_q\dot q_c\dot\xi 	
		+\frac{b_q}{2} q_c(q_c^2-1)\xi+
		\frac{c}{2}q_c(p_c^2-1)\xi
		\right],\nonumber
		\\
		S_2 &=& 
		\frac{1}{2}\int dt
		\left[ a_p\dot \eta^2 + \frac{b_p}{2}(3p_c^2-1)\eta^2
		+\frac{c}{2}(q_c^2-1)\eta^2 + a_q\dot\xi^2
		+\frac{b_q}{2}(3q_c^2-1)\xi^2 \right.\nonumber\\
		&&\left.+\frac{c}{2}(p_c^2-1)\xi^2
		+2cp_c q_c\eta\xi \right], \nonumber\\
		&\equiv& \frac{1}{2}\int dt \Omega^T\mathcal{M}\Omega.
		\label{S2full}
	\end{eqnarray}
	
	The fluctuation vector $\Omega$ is defined as, 
	\begin{equation}
		\Omega= 
		\begin{bmatrix} \eta(t) \\[1ex] \xi(t) \end{bmatrix},\;
		\mathcal{M}= AM_D+M',\;A=\begin{bmatrix} a_p & 0 \\[1ex] 0&  a_q \end{bmatrix}.
	\end{equation}
	$M_D,M'$ are differential operators which are hermitian since the functions on which they operate vanish at the boundaries, 
	\begin{eqnarray}
		M_D&=&\begin{bmatrix} M_{pp} & 0 \\[1ex] 0&  M_{qq} \end{bmatrix}, \quad
		M'=\begin{bmatrix} 0 & M_{pq} \\[1ex] M_{qp}& 0 \end{bmatrix}, \label{M1} \\
		M_{pp}&=&-\frac{d^2}{dt^2}+ V_{pp},\quad
		M_{qq}=-\frac{d^2}{dt^2}+V_{qq},\quad M_{pq}= M_{qp}=V_{pq}.
		\label{M2}
	\end{eqnarray}	The elements of the Hessian matrix  $V_{pp},V_{qq},V_{pq}$ and the dimensionless couplings $\mu,\nu$ are defined below: 
	\begin{eqnarray}
		V_{pp} &=& \frac{\omega_p^2}{2}(3p_c^2 - 1) + \mu\omega_p^2(q_c^2 - 1),\;
		V_{qq} = \frac{\omega_q^2}{2}(3q_c^2 - 1) + \nu\omega_q^2(p_c^2 - 1),\nonumber\\
		V_{pq}&=&V_{qp}= cp_c q_c, \text{ where }
		\mu=\frac{c}{2b_p},
		\quad
		\nu=\frac{c}{2b_q}.
	\end{eqnarray}
	From the definitions in Eq.~\ref{abc} we have $\omega_p^2=b_p/a_p$ and $\omega_q^2=b_q/a_q.$
	The requirement $S_1=0$, then yields the coupled 
	Euler-Lagrange equations for $p_c$ and $q_c$, 
	\begin{eqnarray}
		\frac{d^2p_c}{dt^2} =
		\omega_p^2\left[\frac{1}{2}(p_c^2-1)
		+\mu(q_c^2-1)\right]p_c,\quad
		\frac{d^2q_c}{dt^2} =
		\omega_q^2\left[\frac{1}{2}(q_c^2-1)+\nu(p_c^2-1) \right]q_c.
		\label{EOMp}
	\end{eqnarray}
	
	Insight into this system can be obtained by noting that the classical Euclidean energy $E\equiv -T+V$ is conserved at the classical (but not quantum) level.
	Since the kinetic energy in imaginary time is negative, $E$ becomes: 
	\begin{equation}
		E=-\frac{1}{2}a_p\dot p_c^2 -\frac{1}{2}a_q\dot q_c^2+\frac{1}{8}b_p
		(p_c^2-1)^2+
		\frac{1}{8}b_q(q_c^2-1)^2+\frac{1}{4}c(p_c^2-1)(q_c^2-1).
		\label{energy}
	\end{equation}
	We may now take the derivative and use the EOMs to conclude that $E$ is a constant of motion, 
	\begin{equation}
		\frac{dE}{dt}=
		\frac{d}{dt}(T-V)=0.
		\label{econ}
	\end{equation}
	Since $\dot p_c=0,\dot q_c=0$ when $p,q$ are perched at a hilltop, the kinetic and potential energies both vanish there and so one may choose $E=0$ or, equivalently, $T=V$.
	From Eq.~\ref{S2full} we get a convenient form for the classical action, 
	\begin{eqnarray}
		S_0=\int_{-\frac{T}{2}}^{\frac{T}{2}}
		dt[a_p \dot p_c^2(t)+a_q \dot q_c^2(t)]=S_{0p}+S_{0q},
		\quad
		S_{0p}\equiv a_p||\dot p_c||^2, \;
		S_{0q}\equiv a_q||\dot q_c||^2.
		\label{SS}
	\end{eqnarray}
	The $L_2$ norms used above are defined in the standard way, 
	\begin{eqnarray}
		||f||^2=\int_{-\frac{T}{2}}^{\frac{T}{2}} dt\;f^*(t)f(t).\label{TT}
	\end{eqnarray}
	Energy conservation holds because there is no explicit dependence on time $t$ in the potential $V(p,q)$ in Eq.~\ref{Pot}.
	Along this path the action stays constant, i.e. unless $c=0$ there is a single collective coordinate or single zero eigenvalue mode.
	
	\section{The Zero Mode}
	For uncoupled instantons each instanton has its own zero mode but even the slightest coupling between them results in collapse to a single mode.
	Let us extend the textbook analysis to two coupled instantons. This, as we shall see, is far from trivial.
	In fact we shall be able to achieve an analytic solution only for special cases.
	To obtain the eigenfunction for the soft direction, consider solutions $p_c,q_c$ of the EOM for displaced solutions, $p_c(t,t_c),q_c(t,t_c)$ centred at $t=t_c$.
	This origin is arbitrary and so shifting it by any amount leaves the action unchanged.
	Suppose the shift is by an infinitesimal amount $\delta t_c$, i.e. $t_c \rightarrow t_c+\delta t_c$.
	Then, 	
	\begin{equation}
		S[p_c(t,t_c+\delta t_c),q_c(t,t_c+\delta t_c)] =S[p_c(t,t_c),q_c(t,t_c)]
	\end{equation}
	Using the chain rule, the above condition can be expanded out:
	\begin{equation}
		\begin{aligned}
			0 &= \delta t_c \int dt \left[ \frac{\delta S}{\delta p_c(t)} \dot{p}_c(t) + \frac{\delta S}{\delta q_c(t)} \dot{q}_c(t) \right] \\		&+ \frac{\delta t_c^2}{2} \iint dt \, dt' \left[ \dot{p}_c(t) \frac{\delta^2 S}{\delta p_c(t)\delta p_c(t')} \dot{p}_c(t') + \dot{q}_c(t) \frac{\delta^2 S}{\delta q_c(t)\delta q_c(t')} \dot{q}_c(t') \right.
			\\
			&\quad \left. + 2 \dot{p}_c(t) \frac{\delta^2 S}{\delta p_c(t)\delta q_c(t')} \dot{q}_c(t') \right] \label{SEX}	\end{aligned}
	\end{equation}
	The first term on the RHS above is zero because the first order variations have been required to vanish.
	The second-order term represents the energy cost of shifting the ``center'' $t_c$ of the instanton solution. Because the action is invariant under a global time translation $t \to t + \delta t_c$, the total first and second-order variations with respect to $t_c$ are zero. Eq.~\ref{SEX}, expressed in terms of $\mathcal{M}$ and $\Phi_c$, is 
	\begin{equation}
		\mathcal{M}\dot
		\Phi_c(t)=0,\quad
		\dot\Phi_c\equiv
		\begin{bmatrix}  \dot p_c(t) \\[1ex] \dot q_c(t) \end{bmatrix}.
		\label{zeromode}
	\end{equation}
	Thus $\dot\Phi_c(t)$ is the (unnormalized) zero mode of the two instanton system. This can be reconfirmed by differentiating the EOM's in Eqs.~\ref{EOMp}. 
	The structure of $\mathcal{M}$ above suggests that we consider the generalized eigenvalue problem with $(\mathcal{M},A)$ as the matrix `pencil'  \cite{GEV}, $\mathcal{M}\Phi_n=\lambda_nA\Phi_n,$ which leads to the normalization condition $\int dt\; \Phi_n^TA\Phi_m(t)=\delta_{nm}.$
	
	\section{L-T Decomposition}
\begin{figure}[H]		
		\centering
		\includegraphics[scale=.85]{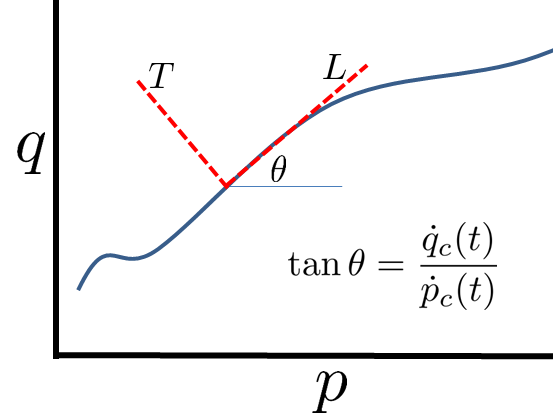}
		\caption{Decomposition of the fluctuation matrix $\mathcal{M}$ in the frame comoving with the instanton.
			The longitudinal axis $L$ represents the soft direction associated with the translational zero mode, while the transverse axis $T$ tracks fluctuations subject to the curvature-induced restoring forces and non-inertial effects in the rotating frame.}
		\label{path}
	\end{figure}

	The coupled differential equations Eqs.~\ref{EOMp} define a path in $(p,q)$ space akin to motion along the bottom of a valley that twists and turns from start to finish.
	The fact that $\mathcal{M}\dot \Phi_c(t)=0$ tells us physically that at every point on the instanton's trajectory there is a soft direction;
	along it a fluctuation can propagate freely whereas to travel perpendicularly requires it to work against a restoring force.
	Intuitively, stability is assured if there is stiffness against a transverse perturbation.
	With this in mind, let us define a frame (see Fig.~\ref{path}) comoving with the instanton and perform an instantaneous, time dependent rotation by $\theta$ with $\theta(t)$ chosen so that the longitudinal axis lies along the soft direction, 
	\begin{eqnarray}
		R=\begin{bmatrix} \cos \theta & -\sin \theta \\[0.2cm] \sin \theta & \cos \theta \end{bmatrix}, \quad 
		\tan\theta=\frac{\dot q_c}{\dot p_c}.
	\end{eqnarray}
	With  $\mathcal{M}\rightarrow \mathcal{M}_{rot}=R^T \mathcal{M} R$, after some considerable algebra we obtain its components in the rotated frame: 
	
	\begin{equation}
		\mathcal{M}_{\text{rot}} = \begin{bmatrix} 		\mathcal{M}_L & \mathcal{M}_{LT}
			\\[0.2cm] 		\mathcal{M}_{TL} & \mathcal{M}_T 
		\end{bmatrix} ,
		\label{Mrot}
	\end{equation}
	where,
	\begin{align}
		\mathcal{M}_L &= \frac{1}{v^2} \left[ (a_p\dot{p}_c^2 + a_q\dot{q}_c^2) \hat{K} + (a_p - a_q)\dot{p}_c\dot{q}_c (2\dot{\theta}\partial_t + \ddot{\theta}) + a_p\dot{p}_c^2 V_{pp} + a_q\dot{q}_c^2 V_{qq} + 2\dot{p}_c \dot{q}_c V_{pq} \right] \nonumber \\[1.5ex]
		\mathcal{M}_T &= \frac{1}{v^2} \left[ (a_p\dot{q}_c^2 + a_q\dot{p}_c^2) \hat{K} - (a_p - a_q)\dot{p}_c\dot{q}_c 
		(2\dot{\theta}\partial_t + \ddot{\theta}) + a_p\dot{q}_c^2 V_{pp} + a_q\dot{p}_c^2 V_{qq} - 2\dot{p}_c \dot{q}_c V_{pq} \right] \nonumber \\[1.5ex]
		\mathcal{M}_{LT} &= \frac{1}{v^2} \left[ -(a_p - a_q)\dot{p}_c\dot{q}_c \hat{K} + (a_p\dot{p}_c^2 + a_q\dot{q}_c^2) (2\dot{\theta}\partial_t + \ddot{\theta}) + \dot{p}_c \dot{q}_c (a_q V_{qq} - a_p V_{pp}) + (\dot{p}_c^2 - \dot{q}_c^2) V_{pq} \right] \nonumber \\[1.5ex]
		\mathcal{M}_{TL} &= \frac{1}{v^2} \left[ -(a_p - a_q)\dot{p}_c\dot{q}_c \hat{K} - (a_p\dot{q}_c^2 + a_q\dot{p}_c^2) (2\dot{\theta}\partial_t + \ddot{\theta}) + \dot{p}_c \dot{q}_c (a_q V_{qq} - a_p V_{pp}) + (\dot{p}_c^2 - \dot{q}_c^2) V_{pq} \right] \label{MTL}
	\end{align}
	The auxiliary quantities are defined as, 
	\begin{eqnarray}		
		\hat{K}= -\frac{d^2}{dt^2} + \dot{\theta}^2, \quad v^2 = \dot{p}_c^2 + \dot{q}_c^2.
	\end{eqnarray}
	Because $\mathcal{M}$ contains the operator $\frac{d^2}{dt^2}$, the time-dependence of $R(t)$ has generated non-inertial terms: $2\dot{\theta} \frac{d}{dt}$, the Euler term $\ddot{\theta}$, and the centrifugal term, $\dot{\theta}^2$.
	Note that although $\mathcal{M}$ is symmetric, $\mathcal{M}_{rot}$ is not - a consequence of fictitious forces in the rotating frame.
	In geometric terms, for any parameterized curve, the curvature of a path $\Phi_c(t) = (p_c(t), q_c(t))$ in the $p$-$q$ plane is given by the kinematic formula: 
	\begin{equation}
		\kappa(t) = \frac{|\dot{p}_c \ddot{q}_c - \dot{q}_c \ddot{p}_c|}{(\dot{p}_c^2 + \dot{q}_c^2)^{3/2}}.
	\end{equation}
	From the relationship $\tan \theta = \dot{q}_c / \dot{p}_c$, the time derivative of the angle is: 
	\begin{equation}
		\dot{\theta} = \frac{\dot{p}_c \ddot{q}_c - \dot{q}_c \ddot{p}_c}{\dot{p}_c^2 + \dot{q}_c^2} = \kappa(t) \sqrt{\dot{p}_c^2 + \dot{q}_c^2}=\kappa(t) v.
	\end{equation}
	
	\section{Solving the EOM's}
\begin{figure}[H]
		\centering		
		\includegraphics[scale=.55]{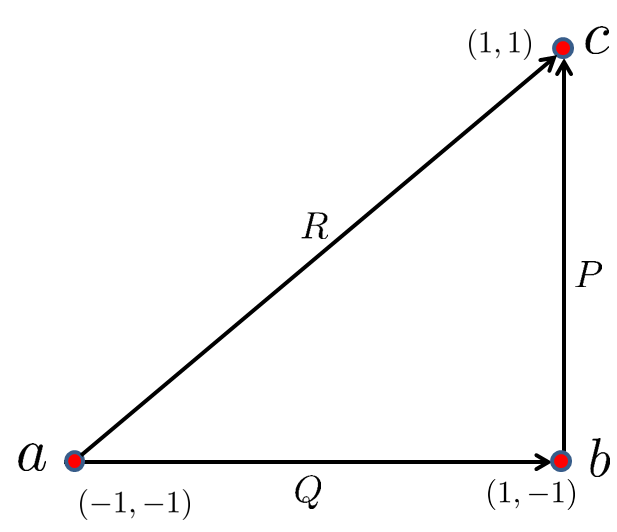}	
		\caption{The three primary instanton configurations in $(p,q)$ space: the edge instantons $P$ and $Q$, which involve a single field transition, and the diagonal $R$ instanton, representing synchronous tunneling of both degrees of freedom.}
		\label{PQR}	
	\end{figure}

	The coupled equations of motion Eqs.~\ref{EOMp} which lead to the classical solutions $p_c(t),q_c(t)$, must now be tackled.
	Various cases are distinguished by the different imposed boundary conditions.
	The null solution $p_c^2(t)=1,q_c^2(t)=1$ satisfies the EOM's with zero action, $S_{0p}=S_{0q}=0$.
	For non-zero action, the solutions fall into 3 types. Two are boundary hugging ($P,Q$) and one is diagonal ($R$): 
	\begin{eqnarray}		
		\textbf{P: } p_c(\mp\infty)&=&\mp 1,\;q_c(\mp\infty)=1
		\\	\textbf{Q: }p_c(\mp\infty)&=&1,\;
		\;q_c(\mp\infty)=\mp 1
		\\
		\textbf{R: }
		p_c(\mp\infty)&=&\mp 1,\;q_c(\mp\infty)=\mp 1
	\end{eqnarray}
	These are illustrated in Fig.~\ref{PQR}. 
	So far our treatment has been both exact and general.
	But solving the coupled non-linear system is a daunting task.
	Let us therefore consider the case where both systems have identical parameters.
	We define the dimensionless quantity $\lambda$, 
	\begin{equation}
		\lambda= m\omega L^2,
	\end{equation} 
	which will serve as the large parameter of the theory.
	With $L$ as the distance between adjacent minima, the system parameters in Eq.~\ref{abc} are re-expressed as, 
	\begin{eqnarray}
		x_p=y_q=L, \quad	\omega_p=\omega_q=\omega,\quad a_p=a_q,\quad \mu=\nu.
	\end{eqnarray}
	With $\tau=\omega t$ as the dimensionless time (henceforth a dot will always denote the derivative with respect to $\tau$), the action takes on a pleasingly simple form, 
	\begin{eqnarray}
		S&=&\lambda\int d\tau
		\bigg[\frac{1}{2}
		\bigg(\dot p^2+\dot q^2\bigg)+V(p,q)\bigg],\nonumber
		\\
		V(p,q)&=&\frac{1}{8}(p^2 - 1)^2 + \frac{1}{8}(q^2 - 1)^2+ 
		\frac{1}{2}\mu(p^2 - 1)(q^2 - 1)
		\label{SSS}
	\end{eqnarray}
	The EOM's are symmetrical: 
	\begin{eqnarray}
		\ddot p_c =\left[\frac{1}{2}(p_c^2-1)
		+\mu(q_c^2-1)\right]p_c,
		\quad \ddot q_c=
		\left[\frac{1}{2}(q_c^2-1)+\mu(p_c^2-1)\right]q_c, \label{EOM1}
	\end{eqnarray} 	
	and naturally suggest looking for a symmetrical solution.
	
	\subsection{Diagonal Instanton} As the Euclidean time goes from $-\infty$ to $\infty$, both $p,q$ move together diagonally from $a$ to $c$ in the form of a single energy lump ($R$ instanton).
	The solution of the EOM's Eq.~\ref{EOM1}, takes the standard tanh form: 
	\begin{eqnarray}
		p_c = q_c = \tanh\frac{\omega_+ \tau}{2},\quad \omega_+ =\sqrt{1+2\mu}.
	\end{eqnarray}
	From Eq.~\ref{SS} the corresponding action is computed to be, 
	\begin{equation}
		S_0^R = \frac{4}{3}
		\lambda \omega_+.
		\label{S0R}
	\end{equation}
	Let us verify that for $p=q$ this is globally the absolute minimum value of the action functional Eq.~\ref{Pot} and satisfies the BPS Saturation (Bogomol'nyi Completion) \cite{BPS1,BPS2} condition for the given boundary conditions, $p(\mp\infty) = \mp 1.$ To this end, rewrite the dimensionless action with $p$ and $q$ set equal to each other, 
	\begin{equation}
		\begin{aligned}
			S=\lambda \int_{-\infty}^{\infty} d\tau\left[\dot{p}^2 + \left(\frac{1}{4}
			+\frac{1}{2}\mu\right)
			(1-p^2)^2 \right] = \lambda
			\int_{-\infty}^{\infty} d\tau \left(\dot{p} - \frac{1}{2}\omega_+(1-p^2) \right)^2 + S_{top}.
		\end{aligned}
	\end{equation}
	The first term is non-negative and is zero only when the first-order BPS equation is satisfied: 
	\begin{equation}
		\dot{p} = \frac{1}{2}\omega_+(1-p^2).
	\end{equation}
	This leaves only $S_{top}$, the topological term defined as, 
	\begin{equation}
		S_{top}= \lambda
		\int_{-\infty}^{\infty} d\tau\dot{p}(1-p^2).
	\end{equation}
	The resulting trajectory is $p_c = \tanh\omega_+ \tau/2$ as in Eq.~\ref{S0R}.
	$S_{top}$, whose value is $S_0^R$ as in Eq.~\ref{S0R}, solely determines the action.
	Further, any path where $p(\tau) \neq q(\tau)$ introduces a non-zero difference field $v(t) = (p-q)/\sqrt{2}$.
	The action functional $S[p, q]$ is minimized when the transverse fluctuations $v(t)$ are zero because the coupling $\mu$ creates a potential valley along the $p=q$ diagonal.
	Hence $S[p \neq q] > S[p = q].$ We shall investigate the transverse stability issue further because this enters critically into calculating the functional determinant.
	
	\subsection{Boundary Instanton, $|\mu| \ll 1$} 
\begin{figure}[H]
\centering
\includegraphics[scale=.40]{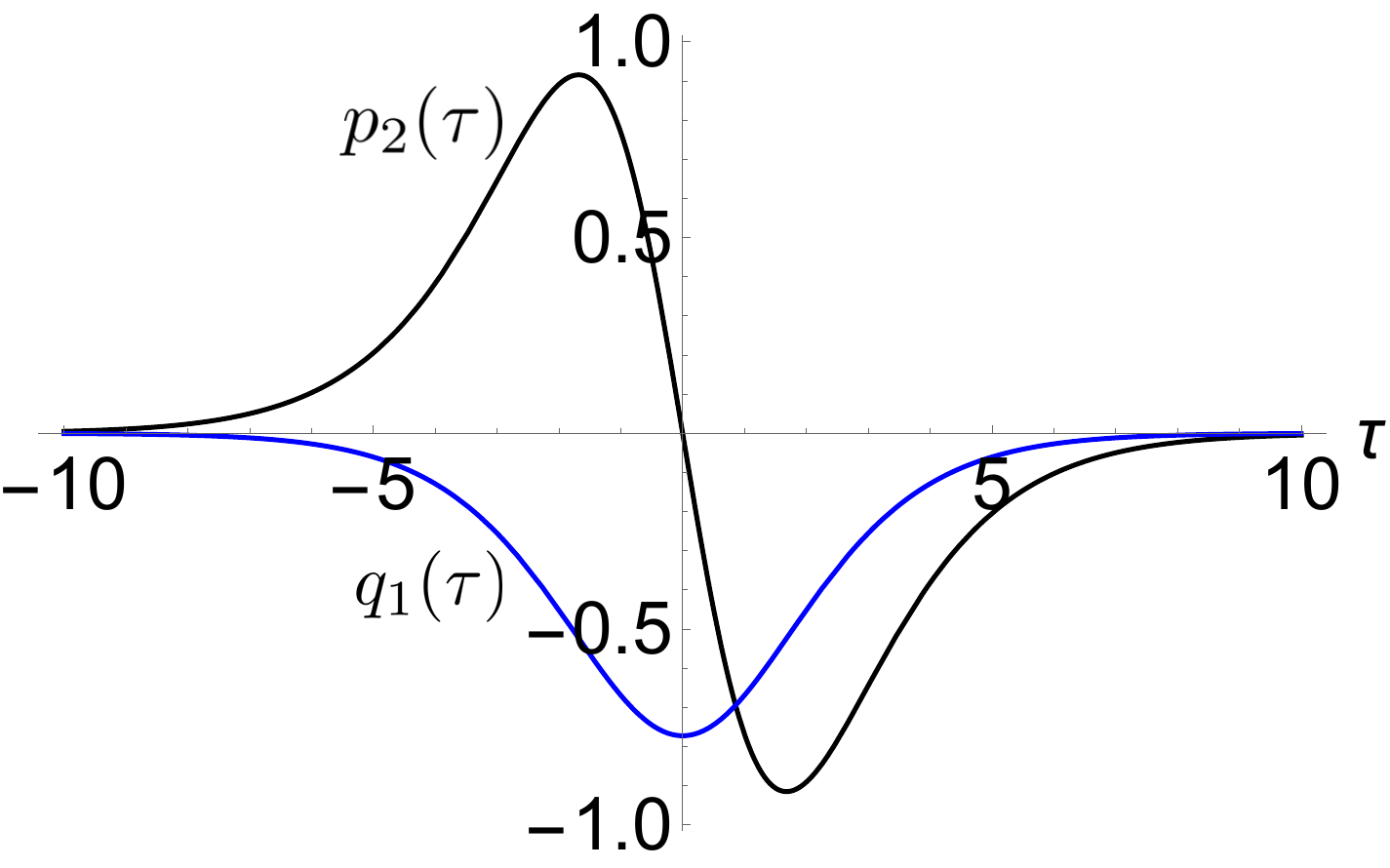}			
\caption{Profiles of the auxiliary functions $q_1(\tau)$ and $p_2(\tau)$ used in the perturbative expansion of the edge instanton action.
The function $q_1(\tau)$ acts as the source for the $p_2$ correction, illustrating the ``dragging'' effect of the secondary field during tunneling.}
\label{q1p2}
\end{figure}

The $P$ instanton essentially hugs the boundary so, as the $p$ field transits to the adjacent minimum, $q$ is dragged along to the origin but eventually returns to its starting position. More specifically: $p_c(\mp \infty) = \mp 1, \quad q_c(\mp\infty) = -1$.
	For arbitrary $\mu$ - unlike the $R$ case - no exact solution is possible and so we must look for a perturbative solution valid for small $\mu$ values, 
	\begin{eqnarray}	p_c(\tau)=p_0+\mu\;
		p_1+\mu^2 p_2+\cdots,\quad
		q_c(\tau)=q_0+\mu\; q_1+\mu^2 q_2+\cdots
		\label{pqex}
	\end{eqnarray}
	The starting point is the uncoupled state, 
	\begin{equation}
		p_0 = \tanh\frac{\tau}{2}, \quad q_0 = -1.
	\end{equation}
	This yields, 
	\begin{equation}
		p_1=0, \text{ and, } \ddot q_1 - q_1 = (p_0^2 - 1)q_0 = \text{sech}^2\frac{ \tau}{2}.
	\end{equation}
	The solution of equation for $q_1$ is,
	\begin{eqnarray}
		q_1 = 2 + 2\tau \sinh\tau- 4\cosh\tau
		\ln\left(2\cosh \frac{\tau}{2}\right).
		\label{q1}
	\end{eqnarray}
	In physical terms, $q_1(\tau)$ is the lump coupled to the $p$ (driving) instanton.
	The source for $p_2(\tau)$ is the interaction term $2\mu q_0 q_1 p_0$: 
	\begin{eqnarray}	\left[ \frac{d^2}{d\tau^2} - \left( 1 - \frac{3}{2}\text{sech}^2 \frac{\tau}{2} \right) \right] p_2(\tau)= -2 \tanh \frac{\tau}{2}\;q_1(\tau).
		\label{pdrive}
	\end{eqnarray}
	An analytical solution for $p_2(\tau)$ is also possible but is non-transparent, requiring a mix of hypergeometric functions.
	Instead, we display the plots of $q_1(\tau)$ and $p_2(\tau)$ in Fig.~\ref{q1p2}.
	Fortunately, by manipulating the driving equation Eq.~\ref{pdrive} for $p_2$ only, we are able to get the kinetic energy to $O(\mu^2)$ in a clean, analytical form, 
	\begin{eqnarray}
		\int_{-\infty}^{\infty} d\tau\;\dot{q}_c^2 \,  = \mu^2 \left( 4\pi^2 - 20 - 16\zeta(3) \right),\quad
		\int_{-\infty}^{\infty} d\tau\;\dot{p}_c^2 = \frac{2}{3} + 16\mu^2 \left[ 2 - \frac{\pi^2}{3} + \zeta(3) \right].
	\end{eqnarray}
	This yields the action to second order, 
	\begin{eqnarray}
		S_0^P=S_0^Q=\frac{2}{3}
		\lambda\left(1-2(\pi^2-9)
		\mu^2+\cdots\right)
		\approx\frac{2}{3}
		\lambda\left(1-1.739
		\mu^2 + \cdots\right).
	\end{eqnarray}
	As can be seen from Fig.~\ref{q1p2}, the energy density of both $p$ and $q$ fields peaks at $t=0$ and rapidly dies away. 
	Therefore this can justifiably be called a 2-D instanton or kink. 
	
	To summarize this section: at the purely classical level, the amplitude for each path in $(p,q)$ space that connects the initial vacuum to the final vacuum is weighted by the factor $\exp(-S_0)$.
	This dominates the tunneling amplitude. For the two specific paths explored, these factors (cf: Fig.~\ref{PQR}) are: 
	\begin{itemize}
		\item 
		a $\rightarrow$
		c:\quad
		$e^{-\frac{4}{3}\lambda \sqrt{1 + 2\mu}}$
		\item 
		a $\rightarrow$ b:\quad
		$e^{-\frac{2}{3}\lambda
			\left(1-2(\pi^2-9)\mu^2+\cdots\right )}$
	\end{itemize}
	For $|\mu|\sim 0$ it is clear that the amplitude for tunneling out to an adjacent minimum $b,d$ is much higher than to the diagonally opposite one, $c$. This will not be the case for larger $|\mu|$ as we shall see shortly.
	
	\subsection{Boundary Instanton, $\mu \sim -\frac{1}{2}$}
	
	With $\mu= -\frac{1}{2}$, the potential in Eq.~\ref{SSS} collapses into,
	\begin{equation}
		V(p,q)\bigg|_{\mu=-\frac{1}{2}}=\frac{1}{8}(p^2-q^2)^2= \frac{1}{2}x^2 y^2.
		\label{coll}
	\end{equation}
	The rotated coordinates are $x=(p+q)/\sqrt{2}$ and $y=(p-q)/\sqrt{2}$.  One observes that the particle motion must be predominantly either along the $x$-axis or $y$-axis and that
	there are no well-defined, isolated vacua. Instead the degenerate vacuum manifold is non-compact, consisting of intersecting continuous valleys. Consequently, standard finite-action instanton solutions connecting discrete minima become ill-defined as $\mu\rightarrow -\frac{1}{2}$. Nevertheless, note that there is no symmetry change in this limit, i.e. the collapsed potential Eq.~\ref{coll} retains the discrete $D_4$ symmetry of the full potential: it is manifestly invariant under parity operations: $p \rightarrow -p$ and $q \rightarrow -q$ as well as the exchange operation: 
	$p\rightarrow q$. (This will not be true in the $\mu\rightarrow \frac{1}{2}$ case, to be dealt with next.)
	
	In the transformed coordinates the EOMs in Eqs.~\ref{EOM1} are:
	\begin{align}
		\ddot{x} &= \frac{\epsilon}{2}x(x^2+y^2-2) + xy^2\approx 
		\frac{\epsilon}{2}x(x^2-2)\\
		\ddot{y} &= \frac{\epsilon}{2}y(x^2+y^2-2) + yx^2
		\approx \frac{\epsilon}{2}y(y^2-2).
	\end{align}
	These must be solved subject to the boundary conditions:
	\begin{align}
		\tau \to -\infty &: (p,q) = (-1,-1) \implies (x,y) = (-\sqrt{2},0) \\
		\tau \to +\infty &: (p,q) = (1,-1) \implies (x,y) = (0,\sqrt{2}).
	\end{align}
	The solution is a two-step process:
	\begin{eqnarray}
		\tau<0: \quad x_c(\tau) &\approx& \sqrt{2}\tanh\left(\sqrt{\frac{\epsilon}{2}} \tau\right), \quad 
		y_c(\tau) \approx 0	\label{xc1}\\	\tau>0: \quad y_c(\tau) &\approx& \sqrt{2}\tanh\left(\sqrt{\frac{\epsilon}{2}} \tau\right), \quad x_c(\tau) \approx 0.
		\label{yc1}
	\end{eqnarray}
	Assuming that the region away from $\tau=0$ can be ignored, the action is trivially calculated,
	\begin{equation}
		S_0^P\bigg|_{\mu\rightarrow -\frac{1}{2}+\epsilon} = \lambda\frac{4\sqrt{2\epsilon}}{3}.
		\label{S0PA}
	\end{equation}
	Observe that this action is exactly that of the $R$ case, Eq.~\ref{S0R}. 
	Also, it is evident that we are seeing the instanton grow wider, and ultimately the total breakdown of the semi-classical approximation. For fixed $\lambda$, going close enough to $\mu=-\frac{1}{2}$ causes a collapse of its fundamental premise that $S_0^P \gg 1$. Further, inspecting Eqs.~\ref{xc1}-\ref{yc1} for $x_c(\tau),y_c(\tau),$ one sees that they are indeed continuous at $\tau=0$ but their derivatives do not exist there. This is a serious matter since the action becomes undefined. We will therefore need a careful scaling argument to see why the central region's contribution is negligible for small $\epsilon$. 
	
	To this end let us focus around the point where the instanton sharply rounds the corner to switch axes. In this region, $x \approx 0$ and $y \approx 0$. The EOMs reduce to:
	\begin{equation}
		\ddot{x} \approx -\epsilon x + xy^2, 
		\quad
		\ddot{y} \approx -\epsilon  y + yx^2.
	\end{equation}
	Let us rescale the fields and the time,
	\begin{equation}
		x = \epsilon^{1/4} X, \quad y = \epsilon^{1/4} Y, \quad \tau = \epsilon^{-1/4} s.
	\end{equation}
	The effective Lagrangian in the transition region is,
	\begin{equation}
		L \approx \epsilon \left[ \frac{1}{2}(X'^2 + Y'^2) + \frac{1}{2}X^2Y^2 + \frac{1}{2} \right],
	\end{equation} 
	which yield the EOMs,
	\begin{equation}
		X''= X Y^2, \quad Y'' = Y X^2.
		\label{newL}
	\end{equation}
	These non-linear equations have no closed form solution else we could have had an analytic form by suitably matching the incoming and outgoing solutions. Nevertheless, we learn from the above that in principle there is a perfectly smooth, differentiable path rather than an abrupt $90^\circ$ turn at the origin. Note that the leading order action near $\tau=0$ is proportional to $\lambda \epsilon$, whereas the action in the large $|\tau|$ domain Eq.~\ref{S0PA} is proportional to $\lambda \sqrt{\epsilon}$. 
	We can offer yet another argument for the suppressed contribution. Near the origin, \begin{equation}
		\ddot{x} \approx -2\epsilon x, \quad \ddot{y} \approx -2\epsilon y. 
	\end{equation}
	The solutions are,
	\begin{equation}
		x(\tau) = A \cos(\sqrt{2\epsilon}\tau +\alpha),\quad	
		y_{in}(\tau)= B \cos(\sqrt{2\epsilon} \tau + \beta).
	\end{equation}
	The kinetic energy contribution to the action from the transition region is clearly $O(\epsilon)$, as was concluded above as well. Unfortunately, calculating the next to leading correction appears to call for numerical work, which we have studiously avoided in this paper.
	
	Still, even at leading order, we have derived an interesting lesson. Although the diagonal transition is sharply suppressed at $\mu=0$, for $\mu\rightarrow -\frac{1}{2}^+$ it has exactly the same action as the boundary hugging instanton and so, at this level of accuracy both paths appear to be equally probable. This may not be actually true for reasons to be discussed later.   
	
	\vspace{0.2cm}
	\subsection{$\mu\rightarrow \frac{1}{2}^+$ case}
	Putting $\mu=\frac{1}{2}$ in the potential Eq.~\ref{SSS} makes it fully $O(2)$ symmetric:
	\begin{equation}
		V(p,q)\bigg|_{\mu= \frac{1}{2}} = \frac{1}{8}\left(p^2 + q^2 - 2\right)^2. 
	\end{equation}
	The four isolated minima of the quartic potential belonging to discrete $D_4$ symmetry are now continuously distributed and the potential barriers have melted away. Hence instanton tunneling ultimately loses meaning. One must now deal with an additional Goldstone mode along the $S^1$ vacuum manifold, augmenting the standard translational zero mode associated with the breaking of Euclidean time-translation invariance. 
	
	Changing to polar coordinates $p=r\cos\theta, q=r\sin\theta$, we now seek minima of the action, i.e. where the discrete minima have not completely dissolved away. To this end let us put $\mu=\frac{1}{2}-\epsilon$ and expand the potential up to $O(\epsilon)$:
	\begin{equation}
		V(r,\theta)\bigg|_{\mu=\frac{1}{2}-\epsilon} = \frac{1}{8}(r^2 - 2)^2 - \frac{1}{2}\epsilon \left( \frac{1}{4} r^4 \sin^2 2\theta - r^2 + 1 \right).
	\end{equation}
	To find the classical instanton path, consider small radial displacements of the form, 
	\begin{equation}
		r^2(\tau) = 2 - \epsilon f(\theta)
	\end{equation}
	Since the zero-energy instanton velocity scales as $\dot{\theta} \sim \mathcal{O}(\epsilon^{1/2})$, the radial velocity scales as $\dot{r} \sim \mathcal{O}(\epsilon^{3/2})$. Substituting the above ansatz into the Lagrangian $\mathcal{L}_E = \frac{1}{2}\dot{r}^2 + \frac{1}{2}r^2 \dot{\theta}^2 + V(r, \theta)$ and expanding in powers of $\epsilon$ gives:
	\begin{align}
		\mathcal{L}_E= \dot{\theta}^2 + \frac{1}{2}\epsilon \cos^2 2\theta
		+O(\epsilon^2)
	\end{align}
	Remarkably, the function $f(\theta)$ has dropped out entirely at leading order. The condition $E=T-V=0$ (see Eq.~\ref{econ}) gives,
	\begin{equation}
		\dot{\theta}^2 = \frac{1}{2}\epsilon \cos^2 2\theta. \label{tdot}
	\end{equation}
	The four potential peaks are located at $\theta=\pm \frac{\pi}{4}, \pm\frac{3\pi}{4}$. 
	With boundary conditions $\theta(-\infty) = -\frac{3\pi}{4}$ and $\theta(+\infty) = -\frac{\pi}{4}$, integrating Eq.~\ref{tdot} gives:
	\begin{equation}
		\theta_c(\tau) = -\frac{\pi}{2} + \frac{1}{2} \arcsin\left(\tanh(\sqrt{2\epsilon}\tau)
		\right).\label{tc}
	\end{equation}
	The instanton action $S_0$ evaluates to:
	\begin{equation}
		S_0\bigg|_{\mu=\frac{1}{2}-\epsilon} = 2\lambda \int_{-3\pi/4}^{-\pi/4} d\theta \sqrt{\frac{\epsilon}{2}} \left| \cos 2\theta \right| = \lambda \sqrt{2\epsilon}.
	\end{equation}
	This is almost identical to that at the left edge, Eq.~\ref{S0PA}.
	
	\section{ Fluctuation Operators}
	With solutions now in hand for the EOM's - exact for the diagonal case and approximate for the edge instantons - we proceed to the next stage, i.e. the quadratic level fluctuations around the classical solutions. These lead to pre-factors multiplying the exponential of the action. In every case below we shall summarize the stability criteria and spectrum of the fluctuation operators, most of which are of the P\"oschl-Teller type, the treatment of which is spread over several classic textbooks \cite{Landau, Flugge, Kleinert}.
	
	\subsection{Stability and Spectrum}
	The fluctuation operators encountered in the remainder of this section are of the type,
	\begin{equation}
		\hat M = -\frac{d^2}{d\tau^2} + m^2 - J^2\text{sech}^2\Omega \tau = \Omega^2 \hat{O}[\kappa,j]
		\label{Posch}
	\end{equation}
	The eigenvalue problem $\hat M\psi = \lambda\psi$ takes the form,
	\begin{equation}
		\left[ -\frac{d^2}{dx^2} - \frac{j(j+1)}{\Omega^2}\text{sech}^2 x \right] \psi(x) = \frac{\lambda - m^2}{\Omega^2} \psi(x)
	\end{equation}
	To identify with the standard form of the P\"oschl-Teller operator, 
	\begin{equation}
		\hat{O}[\kappa,j]=-\frac{d^2}{dx^2} + \kappa^2 - j(j+1)\text{sech}^2x,
	\end{equation}
	we had to put,  
	\begin{equation}
		x=\Omega\tau,\quad 
		\kappa^2=\frac{m^2}	{\Omega^2},\quad j(j+1) = \frac{J^2}{\Omega^2}.
		\label{param}
	\end{equation}
	The discrete eigenvalues for $H = -\partial_x^2 - j(j+1)\text{sech}^2x$ are known to be $\mathcal{E}_n = -(j - n)^2$, where $n = 0, 1, 2, \dots$ strictly bounds $n <j$. Hence the discrete spectrum of $M$ is:
	\begin{equation}
		\lambda_n =\Omega^2[ \kappa^2 - (j - n)^2].	
	\end{equation}
	For $M$ to be stable against perturbations, it cannot possess any negative eigenvalues. Thus the stability criterion is, $\lambda_n \ge 0$ for all valid $n$ or,
	\begin{equation}
		\lambda_0 =\Omega^2( \kappa^2 - j^2)
		\ge 0. \label{lam0}
	\end{equation}
	
	Next, we must find the functional determinant for $\hat M$ in Eq.~\ref{Posch}. 
	The Gelfand-Yaglom \cite{Gelfand} method - reviewed by Dunne \cite{Dunne}, Casahorran \cite{Casahorran}, and others is ideally suited for this purpose. Instead of evaluating an infinite product of eigenvalues, one need only solve an initial value problem on the interval $[-T, T]$ with Dirichlet boundary conditions. The ratio of determinants is then calculated from the asymptotic limit: 
	\begin{equation}
		R(\lambda) = \frac{\det \hat{O}[\kappa,j]}
		{\det \hat{O}[\kappa,0]} = \lim_{T \to \infty} \frac{\phi(T)}{\phi_0(T)}.
	\end{equation}
	Here $\hat{O}[\kappa,0]$ is the reference operator. With $\phi(t)$ as the solution to the initial value problem under the differential operator $\hat{O}[\kappa,j]$, the condition is imposed on the left boundary $x = -T$ is:
	\begin{equation}
		\phi(-T) = 0, \quad \phi'(-T) = 1.
	\end{equation}
	The reference operator is subject to exactly the same initial conditions:
	\begin{equation}
		\phi_0(-T) = 0, \quad \phi_0'(-T) = 1.
	\end{equation}
	A closed form result in terms of Gamma functions is obtained:
	\begin{equation}
		R= \frac{\Gamma\left(\kappa + 1\right) \Gamma\left(\kappa
			\right)}{\Gamma\left(\kappa - j\right) \Gamma\left(\kappa + j + 1\right)}. 
		\label{ratio}
	\end{equation}
	We shall now apply this theorem to both the diagonal and edge cases.
	\subsection{Diagonal Case}
	Having an exact solution of the EOMs enables us to completely analyze the diagonal case. As per our earlier discussion, the $p,q$ instantons travel together along the diagonal $\theta = \pi/4$ in $p,q$ space with $\dot{p}_c = \dot{q}_c$ and $\dot{\theta}= \ddot{\theta} = 0.$ 
	Linear motion eliminates all non-inertial couplings and off-diagonal terms in $\mathcal{M}_{rot}$, Eq.~\ref{Mrot}.
	With $p_c = q_c = \tanh\omega_+ \tau/2$, the potentials are: 	\begin{align}		V_{pp} = V_{qq}= \frac{1}{2}(3p_c^2 - 1) + \mu(p_c^2 - 1),\quad	
		V_{pq} = c p_c^2 = 2 \mu p_c^2.
	\end{align}
	The rotated matrix $\mathcal{M}_{rot}$ becomes exactly diagonal: 	\begin{eqnarray}
		\mathcal{M}_{rot}^R =\lambda \begin{bmatrix} \mathcal{M}_L^R & 0\\[0.2cm] 0 & \mathcal{M}_T^R \end{bmatrix}.\label{MR}
	\end{eqnarray}		
	The longitudinal operator $\mathcal{M}_{L}^{R}$, and the transverse operator $\mathcal{M}_T^R$, reduce to: 
	\begin{eqnarray}	\mathcal{M}_{L}^{R} &=& -\frac{d^{2}}{d\tau^{2}} +\omega_+^2 -\frac{3}{2} \omega_+^2\text{sech}^{2}
		\frac{\omega_+ \tau}{2},
		\label{OTL} \\		\mathcal{M}_T^R &=& -\frac{d^2}{d\tau^2} + \omega_-^2 -\left(\frac{3}{2}
		-\mu
		\right)
		\text{sech}^2\frac{\omega_+ \tau}{2},
		\nonumber\\
		\text{where }
		\omega_\pm^2&=&1\pm 2\mu.
		\label{OTT}
	\end{eqnarray}
	As discussed in Appendix A, $\omega_\pm$ are the scaled frequencies of the two independent oscillator modes that follow from diagonalizing the Hamiltonian near any of the four potential minima.
	
	We next turn to stability issues. Because the lowest eigenvalue of $\mathcal{M}_{L}^{R}$ is exactly zero (representing the translational zero mode) and all other eigenvalues are strictly positive, there are no runaway solutions. Thus the configuration is stable against small fluctuations. The transverse operator $\mathcal{M}_T^R$ demands closer inspection. Stability against perturbations is determined by the lowest eigenvalue $\lambda_0$ of the spectral problem $\mathcal{M}_T^R\psi = \lambda\psi$. This evaluates to,
	\begin{equation}
		\lambda_0=\frac{1}{2} \sqrt{\frac{25-14 \mu }{2 \mu +1}}-\frac{5}{2}
	\end{equation}
	There are four distinct regimes corresponding to all possible values of $\mu$:
	
	\begin{itemize}
		\item
		\textbf{Breakdown Regime ($\mu \le -1/2$):} The frequency $\omega_+$ becomes imaginary; there is no kink. 
		\item \textbf{Attractive/Negative Coupling ($-1/2 < \mu < 0$):} In this regime, $0 < \omega_+^2 < 1$ and $\lambda_0$ evaluates to strictly positive values implying strict stability. The diagonal instanton exists only for an attractive potential between $p,q$ fields.
		
		\item \textbf{Uncoupled Limit ($\mu = 0$):} Here $\lambda_0 =  0$. The operator is marginally stable, reflecting the presence of an unbroken translational zero mode.
		
		\item \textbf{Repulsive/Positive Coupling ($\mu > 0$):} In this regime, $\omega_+^2 > 1$. The operator is strictly unstable.
	\end{itemize}
	
	\noindent The upshot of the above analysis is that we may entirely exclude the diagonal instanton for $\mu>0$. Let us now proceed to calculate the fluctuation operators Eqs.~\ref{OTL}-\ref{OTT} for the diagonal case and $-1/2 < \mu < 0$. 
	
	First for the longitudinal operator: Eq.~\ref{OTL} has, as per Eqs.~\ref{param}, the parameters $\kappa=j=2$. From this one gets $\lambda_0=0$. Thus the longitudinal mode has marginal stability but $\det \mathcal{M}_L^R$ vanishes. This, of course, is the expected zero mode, to remove which one takes the ratio:
	\begin{equation*}
		\chi_L^R =\frac{\det \mathcal{M}_{0L}^R}
		{\det \mathcal{M}_L^R}
		= \lim_{\kappa \to j} \frac{\Omega^2(\kappa^2 - j^2)}{R_L},
	\end{equation*}
	where $R_L$ is from Eq.~\ref{ratio}. 
	Using $z\Gamma(z) = \Gamma(z+1)$ in Eq.~\ref{ratio} for $R$ and $\Omega =\omega_+ /2$, this ``primed" determinant becomes, 
	\begin{equation}
		\chi_L^R =12\omega_+^2.
	\end{equation}
	
\begin{figure}[H]
		\centering
		\includegraphics[scale=.4]{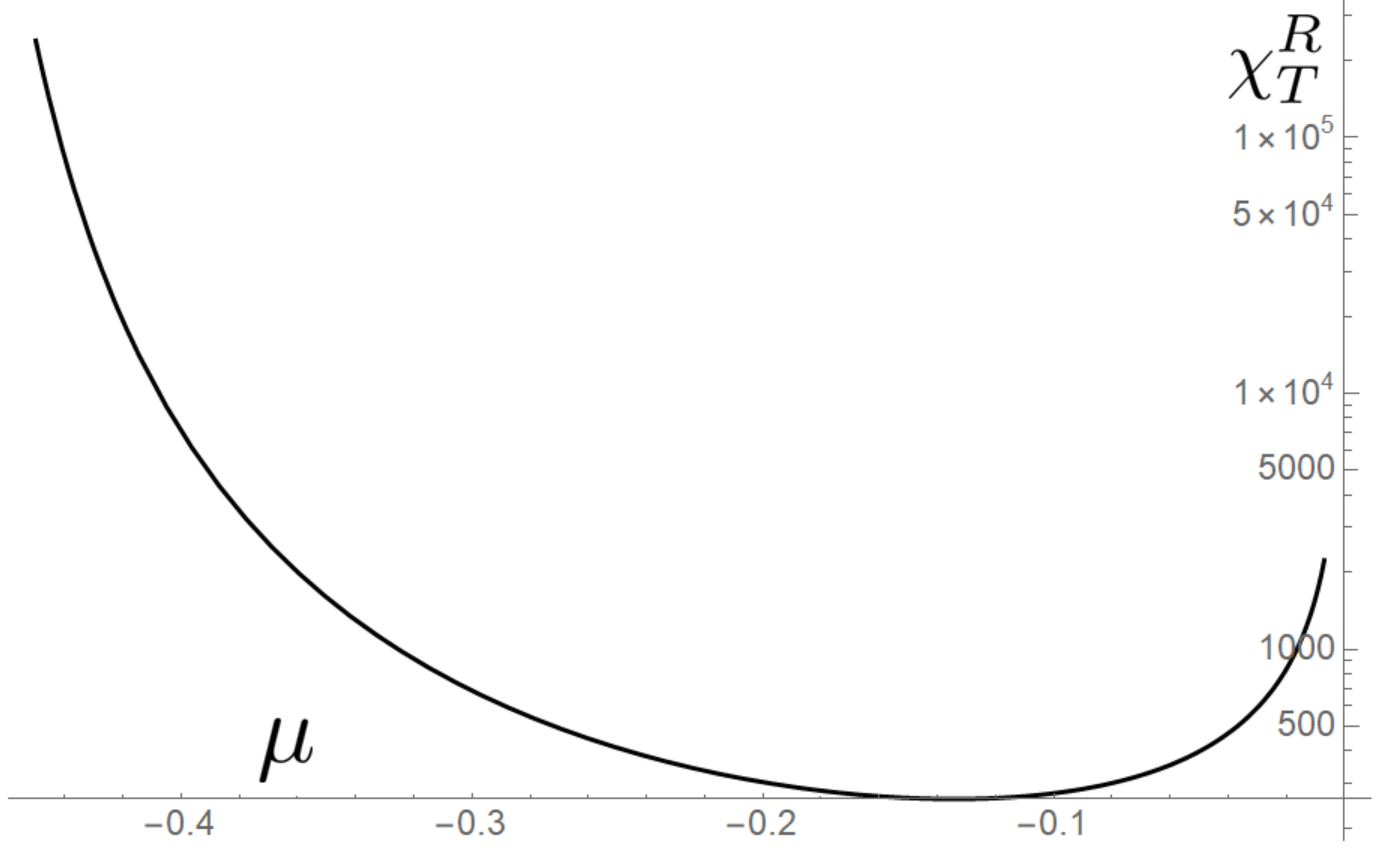}
		\caption{The ratio of determinants $\chi_T^R(\mu)$ (see Eqs.~\ref{chitr}-\ref{chai}) as a function of the coupling constant $\mu$.
			The divergence as $\mu \to 0$ indicates the presence of an additional zero mode in the uncoupled limit, while the behavior near $\mu = -0.5$ signals the loss of discrete minima in the tunneling potential.}
		\label{chi}
	\end{figure}

	Second, for the transverse operator: in this region of $\mu, \lambda_0>0$ ensures stability. The determinant ratio is Eq.~\ref{ratio} is, 
	\begin{eqnarray}
		\chi_T^R(\mu)=\frac{\det \mathcal{M}_{0T}^R}
		{\det \mathcal{M}_T^R}	=\frac{\Gamma(\kappa - j)		\Gamma(\kappa + j+1)}
		{\Gamma(\kappa)		\Gamma(\kappa + 1)},
		\quad \label{chitr}
	\end{eqnarray}
	where $\kappa,j$ are functions of $\mu$,  
	\begin{eqnarray}
		\kappa^2 = \frac{4(1 - 2\mu)}{1 + 2\mu},\quad
		j(j+1) &=& \frac{6 - 4\mu}{1 + 2\mu}.
		\label{chai}
	\end{eqnarray}
	Anticipating that the amplitude for diagonal travel will be proportional to the product $\chi_T^L\chi_T^R $, let us reflect upon the qualitative behavior of $\chi_T^R(\mu)$. This is plotted in Fig.~\ref{chi} and is divergent at both ends, $\mu=0$ and $\mu=-1/2.$
	
	For $\mu=0$ we have $\kappa=\ell=2$ and hence, from Eq.~\ref{chitr} $\chi_T^R(0)\sim\Gamma(0)$, which is infinite.
	This means that the uncoupled fields are, as one expected, free to move both longitudinally and transversely, i.e. there are two zero modes.
	Expanding near $\mu=0$ gives $\chi_T^R(\mu)\sim -15/\mu$. For the amplitude to be real, $\mu$ must therefore be negative, i.e. the potential between $p,q$ must be attractive for them to travel together on a diagonal path.
	The behavior as $\mu \to -1/2^+$ is even more intriguing.
	If we take $\epsilon = \mu + 1/2$ 
	then $\kappa = 2/\sqrt{\epsilon} - \sqrt{\epsilon}$ while $\ell =2/\sqrt{\epsilon} - 1/2$.
	After some further analysis, and using $\log n!=n \log n -n$ (Stirling's  formula) one concludes that an essential singularity exists at $\mu=-1/2$: \begin{equation}
		\chi_T^R \sim  \exp\left( \frac{4\ln 2}{\sqrt{\epsilon}} \right).
		\label{sing}
	\end{equation}
	The physical cause appears upon inspecting the potential: as $\mu$ approaches $-0.5$, the diagonal barrier vanishes, $\omega_+ \to 0$.
	While the discrete $D_4$ spatial symmetry remains intact, no longer are there standard finite-action instanton solutions connecting discrete minima. 
	
	\subsection{Boundary instanton, $|\mu|\ll 1$}
	The rotation matrix $\mathcal{M}_{rot}^P$ for the $P$ system contains both velocity dependent diagonal as well as off-diagonal terms.
	The Coriolis terms make it non-symmetric. 
	\begin{align}			\mathcal{M}_L^P &= -\frac{d^2}{d\tau^2} + \dot{\theta}^2 + \frac{V_{pp}\dot{p}_c^2 + V_{qq}\dot{q}_c^2 + 4\mu V_{pq}\dot{p}_c\dot{q}_c}{\dot{p}_c^2 + \dot{q}_c^2} \\[10pt]
		\mathcal{M}_T^P &= -\frac{d^2}{d\tau^2} + \dot{\theta}^2 + \frac{V_{pp}\dot{q}_c^2 + V_{qq}\dot{p}_c^2 - 4\mu V_{pq}\dot{p}_c\dot{q}_c}{\dot{p}_c^2 + \dot{q}_c^2} \\[10pt]
		\mathcal{M}_{LT}^P &= 2\dot{\theta}\frac{d}{d\tau} + \ddot{\theta} + \frac{(V_{qq} - V_{pp})\dot{p}_c\dot{q}_c + 2\mu V_{pq}(\dot{p}_c^2 - \dot{q}_c^2)}{\dot{p}_c^2 + \dot{q}_c^2} \\[10pt]
		\mathcal{M}_{TL}^P &= -2\dot{\theta}\frac{d}{d\tau} - \ddot{\theta} + \frac{(V_{qq} - V_{pp})\dot{p}_c\dot{q}_c + 2\mu V_{pq}(\dot{p}_c^2 - \dot{q}_c^2)}{\dot{p}_c^2 + \dot{q}_c^2}
	\end{align}
	Given the complexity, it is natural to seek a perturbative expansion of the operators.
	The moving frame rotation angle is set by $\theta=\tan^{-1}\dot q_c/\dot p_c\sim O(\mu)$.
	From the previous section the edge solution is $p_c\approx\tanh \tau/2$ and $q_c\approx -1+\mu q_1$.
	Hence, up to $O(\mu)$, 
	\begin{eqnarray}
		\mathcal{M}_L^P &=& -\frac{d^2}{d\tau^2} + 1 -\frac{3}{2}\text{sech}^2\frac{\tau}{2}
		\label{MLP}\\	\mathcal{M}_T^P  &=& -\frac{d^2}{d\tau^2} + 1 -\mu\left(\text{sech}^2\frac{\tau}{2} + 3q_1\right)\label{MTP} \\
		\mathcal{M}_{LT}^P  &=& 2\dot{\theta}\frac{d}{d\tau} + \ddot{\theta} + 3\mu \dot{q}_1\\
		\mathcal{M}_{TL}^P  &=& -2\dot{\theta}\frac{d}{d\tau} - \ddot{\theta} + 3\mu \dot{q}_1.
	\end{eqnarray}
	If the spectrum of $\mathcal{M}_{rot}^P$ is to considered only up to $O(\mu)$ then it is effectively diagonal and we may set $\mathcal{M}_{TL}^P
	\times  \mathcal{M}_{LT}^P
	\approx 0$.
	
	Let us examine the spectrum of $\mathcal{M}_T^P$ in Eq.~\ref{MTP}. For stability, its lowest eigenvalue must be strictly positive, $\lambda_0 > 0$. The function $ \text{sech}^2\tau/2 + 3q_1(\tau) $ is strictly negative for all values of $\tau$ and so, the effective mass is always positive for repulsive potentials, $\mu>0$. If $\mu < 0$ we 
	can use the weak-coupling approximation:
	\begin{equation}
		\lambda_0 \approx 1 - \frac{1}{4}\left(\int_{-\infty}^{\infty} V \, d\tau\right)^2 = 1 - 16\mu^2
	\end{equation}
	Since we are presently evaluating the negative regime, this restricts the parameter to $-\frac{1}{4} < \mu < 0$.
	
	Let us now turn towards calculating the fluctuation determinants: the zero mode longitudinal operator Eq.~\ref{MLP} is verbatim that of the diagonal instanton (except that here $\omega_+ \rightarrow 1$) and so the primed determinant is,
	\begin{equation}
		\chi_L^P(\mu)=\frac{\det\mathcal{M}_{0L}^P}
		{\det' \mathcal{M}_L^P} = 12.
	\end{equation}
	The transverse operator is not of the P\"oschl-Teller kind and will require additional work. To this end, decompose it in a free operator and a perturbative part:
	\begin{equation}
		\mathcal{M}_T^P \equiv \mathcal{M}_{0T}^P+V = \left(-\frac{d^2}{d\tau^2} + 1\right) - \mu \left[ 3q_1(\tau) + \text{sech}^2\left(\frac{\tau}{2}\right) \right]
	\end{equation}
	To evaluate the ratio of determinants, we recall the trace-log identity, $\ln(\det(A)) = \text{Tr}(\ln(A))$:
	\begin{eqnarray}
		\ln\left( \frac{\det \mathcal{M}_T^P}{\det \mathcal{M}_{0T}^P} \right) &=& \ln\left( \det(\mathcal{M}_{0T}^{P-1}\mathcal{M}_T^P) \right) = \text{Tr}\left( \ln(\mathcal{M}_{0T}^{P-1}\mathcal{M}_T^P) \right)\\
		&=&\text{Tr}\left( \ln(\mathbf{1} + \mathcal{M}_{0T}^{P-1}V) \right) \approx \text{Tr} \left( \mathcal{M}_{0T}^{P-1}V \right)
	\end{eqnarray}
	The trace is evaluated as an integral over the diagonal matrix elements:
	\begin{equation}
		\text{Tr}\left( \mathcal{M}_{0T}^{P-1}V_{pert} \right) = \int_{-\infty}^{\infty} d\tau \, \langle \tau | \mathcal{M}_0^{P-1} | \tau \rangle V(\tau)
	\end{equation}
	The matrix element $\langle \tau | \mathcal{M}_0^{-1} | \tau \rangle$ is the coincidence limit of the free Green's function, $G(\tau, \tau')$, satisfying:
	\begin{equation}
		\left( -\frac{d^2}{d\tau^2} + 1 \right) G(\tau, \tau') = \delta(\tau - \tau'),
	\end{equation}
	the solution to which is:
	$G(\tau, \tau') = \frac{1}{2} e^{-|\tau - \tau'|}$.
	At $\tau = \tau'$ this gives $\mathcal{M}_{0T}^{P-1}=G(\tau, \tau) = \frac{1}{2}$. 
	
	Substituting $G(\tau, \tau)$ and the explicit form of $V$ back into the integral, we obtain the final perturbative expression:
	\begin{equation}
		\ln\left( \frac{\det \mathcal{M}_T^P}{\det\mathcal{M}_{0T}^P}
		\right) \approx \int_{-\infty}^{\infty} d\tau \, \frac{1}{2} \left( -\mu \left[ 3q_1(\tau) + \text{sech}^2\left(\frac{\tau}{2}\right) \right] \right)
	\end{equation}
	To get the first integral above on the RHS, we can integrate both sides of Eq.~\ref{q1},
	$-\ddot{q}_1 + q_1 = -\text{sech}^2\frac{\tau}{2}.$  This gives, 
	\begin{equation}
		\int_{-\infty}^{\infty} q_1(\tau) \, d\tau = -4 \implies \text{RHS}=4. 
	\end{equation}
	So, finally, 
	\begin{equation}
		\chi_T^P(\mu)=\frac{\det\mathcal{M}_{0T}^P}
		{\det \mathcal{M}_T^P} = e^{-4\mu}\approx 1 - 4\mu.
	\end{equation}
	
	\subsection{Edge Case, $\mu\rightarrow -\frac{1}{2}^+$}
	
	Informed by the solution to the EOMs in Eqs.~\ref{xc1}-\ref{yc1} for $\tau<0$, we can set to zero the centrifugal and Coriolis terms in the general formula Eq.~\ref{MTL}. With $\mu=-\frac{1}{2}+\epsilon$, the fluctuation operator $\hat{\mathcal{M}}$ is then purely diagonal with components:
	\begin{eqnarray}
		\hat{\mathcal{M}}_L^P\bigg|_{\mu\rightarrow -\frac{1}{2}^+} (\tau<0)&=& -\partial_\tau^2 + 2\epsilon-3\epsilon \, \text{sech}^2
		\sqrt{\frac{\epsilon}{2}}\tau,
		\label{m11}
		\\
		\hat{\mathcal{M}}_T^P\bigg|_{\mu\rightarrow -\frac{1}{2}^+} (\tau<0)&=& -\partial_\tau^2
		+2(1-\epsilon)
		-(2-\epsilon) \, \text{sech}^2
		\sqrt{\frac{\epsilon}{2}}\tau.
		\label{m12}
	\end{eqnarray}
	The above are precisely the operators for the diagonal case, Eqs.~\ref{OTL}-\ref{OTT}. For $\tau>0$, as per our discussion in Section 5.3, the longitudinal and transverse labels are switched as the instanton takes a $90^\circ$ turn at the origin. This leaves the determinant of the product of the two operators unaffected. But a caveat precludes a proper calculation of the full fluctuation determinant: the $\tau\approx 0$ transition region induces strong mixing through off-diagonal components. We have not succeeded in calculating these analytically although the effective Lagrangian Eq.~\ref{newL} in the transition region suggests that a matching strategy can perhaps be found. We will not pursue this further here. The main conclusion to be drawn from this limiting case - as for the diagonal instanton $R$ - is that the fluctuation determinant dominates as $\mu\rightarrow -\frac{1}{2}^+$, i.e. as the barrier height goes to zero, the semi-classical approximation breaks down. 
	
	\subsection{Edge Case, 
		$\mu\rightarrow \frac{1}{2}$}
	A very different situation awaits us at the opposite edge of the $\mu$ line. Taking the dimensionless action in Eq.~\ref{SSS} as our starting point, switch to polar coordinates $p=r\cos\theta$ and $q=r\sin\theta$, and expand to $O(\epsilon)$, where $\epsilon = \frac{1}{2}-\mu$.
	We now examine small fluctuations around the classical instanton solution, \begin{equation}
		r(\tau) = r_c(\tau) + \rho(\tau),\quad \theta(\tau) = \theta_c(\tau) +
		\eta (\tau) 
	\end{equation}
	Keeping only quadratic level terms in $\rho,\eta$ and using $r_c \approx \sqrt{2}$, the following quadratic Lagrangian emerges:
	\begin{equation}
		\mathcal{L}^{(2)} = \frac{1}{2}\dot{\rho}^2 + \dot{\eta}^2 + \rho^2 \left( 1 + \frac{1}{2}\dot{\theta}_c^2 \right) + 2\sqrt{2}\rho\dot{\theta}_c\dot{\eta} - 2\epsilon \eta^2 \cos 4\theta_c.
	\end{equation}
	In the adiabatic approximation one can neglect the motion of the heavy mode $\rho$, then minimize to find the $\rho$ value. To lowest order, 
	
	\begin{equation}
		\mathcal{L}^{(2)} = \dot{\eta}^2  - 2\epsilon\eta^2\left( 2\text{sech}^2(\sqrt{2\epsilon}\tau) - 1 \right)
		\implies \mathcal{M} = -\frac{d^2}{d\tau^2} + 2\epsilon - 4\epsilon\text{sech}^2(\sqrt{2\epsilon}\tau).
	\end{equation}
	The primed determinant is,
	\begin{equation}
		\chi_T^P(\mu)\bigg|_{\mu\rightarrow \frac{1}{2}}=8\epsilon.
	\end{equation}
	There is no violent behavior of $\chi_T$ here even though the semi-classical approximation (for fixed $\lambda$ ) is breaking down close to this edge as well: $S_0\sim\lambda \sqrt{2\epsilon}$.
	
	\section{Transition Amplitudes}
	Having found the classical paths, actions, and fluctuation determinants, we have almost completed our journey of finding the transition amplitudes. The central formula is derived in Appendix A: $\mathcal{A}=CKT$ where the entire instanton dynamics is encapsulated in the transition rate $K$,
	\begin{equation}
		K=\sqrt{\frac{S_0}{2\pi}}e^{-S_0}	\sqrt{\frac{\det A^{-1}\mathcal{M}_0}{\det' A^{-1}\mathcal{M}_{rot}}}.
	\end{equation}
	The determinant of the fluctuation determinant is symbolically,
	\begin{equation}
		\frac{\det}{\det'}=\chi_L\times \chi_T
	\end{equation}
	All ingredients are now in place for evaluating $K$ in each of the cases we have analyzed. The results can be thus summarized:
	
	\begin{itemize}
		\item 
		Diagonal Instanton, $-\frac{1}{2}<\mu<0$
		\begin{eqnarray}
			S_0 &=& \frac{4}{3} \lambda \omega_+,\quad 
			\chi_L=12\omega_+^2,\quad
			\chi_T=
			\frac{\Gamma(\kappa-j)
				\Gamma(\kappa +j+1)}
			{\Gamma(\kappa)
				\Gamma(\kappa + 1)},
			\nonumber \\
			\omega_+^2&=&1+2\mu,\quad
			\kappa^2 = \frac{4(1 - 2\mu)}{1 + 2\mu},\quad
			j(j+1)=\frac{6-4\mu}{1 + 2\mu}.\label{d1}
		\end{eqnarray}
		\item 
		Boundary Instanton, 
		$|\mu|\ll 1$ 
		\begin{eqnarray}
			S_0=\frac{2}{3}
			\lambda\left(1-2(\pi^2-9)	\mu^2+\cdots\right),\quad \chi_L=12,\quad\chi_T=1-4\mu.
			\label{d2}
		\end{eqnarray}
		\item 
		Boundary Instanton, 
		$\mu=-\frac{1}{2}+\epsilon$ 
		
		The values of $S_0, \chi_L,\chi_T$ are exactly those for the diagonal case. However, 
		$\det/\det'\ne\chi_L\times \chi_T$ (not factorizable). The fluctuation determinant apparently eludes an analytical form because of the transition region near $\tau\sim 0$. See discussion in Section 6.4
		\item  
		Boundary Instanton, 
		$\mu=\frac{1}{2}-\epsilon$
		\begin{equation}
			S_0 =\lambda \sqrt{2\epsilon},\quad 
			\chi_L=8\epsilon,\quad
			\chi_T=1.
			\label{d3}	
		\end{equation}
		
	\end{itemize}
	
	Before proceeding to numerically evaluating $K$, we recall that the semi-classical approximation makes sense only if $\exp(-S_0)$ dominates, i.e. $S_0\gg 1$. For small enough value of the scaling parameter - i.e. large enough $\lambda$ - this translates differently in the four cases above: $\lambda\gg |\mu+\frac{1}{2}|^{-\frac{1}{2}},\; \lambda\gg 1,\; \lambda\gg |\mu+\frac{1}{2}|^{-\frac{1}{2}}, \;\lambda\gg |\mu-\frac{1}{2}|^{-\frac{1}{2}}$. But this may not be a sufficient condition. We have seen that evaluating the fluctuation determinant on the left edge led to an essential singularity. If the classical term is to truly dominate the dynamics, Eq.~\ref{sing} tells us that the requirement is even more stringent on the left edge as compared to the right edge,
	\begin{equation}
		\lambda\gg \bigg|\mu+\frac{1}{2}\bigg|^{-1}.\label{lbig}
	\end{equation}
	In physical terms, the shallowness of the potential trough enhances the quadratic fluctuations and makes them competitive with the classical level amplitude.
	
	Before turning to the next stage of analysis, let us recall that in the 1-D instanton case the contribution of a single instanton to $\mathcal{A}$ vanishes because $Te^{-\omega T}\rightarrow 0$ as $T\rightarrow \infty$.
	But the classical EOMs admit solutions beyond those considered so far. For one, reversing $t$ gives the anti-instanton of that flavor.
	For another, any number of well separated, well localized instantons and anti-instantons is also a solution if it satisfies the boundary conditions. This, of course, leads to the dilute instanton gas approximation (DIGA), to be covered next. The validity of DIGA requires: a)that the single instanton action be sufficiently large, b)the average temporal separation between successive instantons (and anti-instantons) must be much larger than the characteristic width of a single instanton. This width is $\mu$ dependent and is identified from the various EOM solutions in Section 5. In principle, an instanton gas could be made arbitrarily dilute by raising $\lambda$ sufficiently. The bottom line: for fixed $\lambda$, DIGA is guaranteed to fail if we get close enough to either edge of the $\mu$ line. 
	
	\section{Dilute 3-Flavor Gas}
	
	Extending from the single flavor dilute gas model to three flavors is now our goal. It will not be assumed we are dealing with the symmetric case and $K_P,K_Q,K_R$ can be arbitrary in what follows in this section.
	\begin{figure}[H]
centering
\includegraphics[scale=.4]{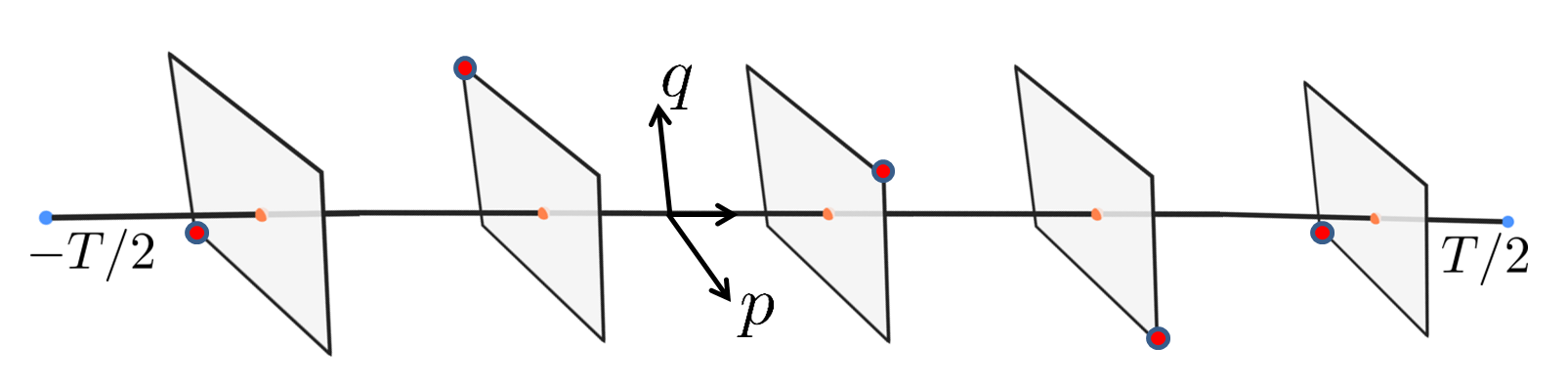}
\caption{A chain of 2 DOF instantons as they journey from start to end.
The edges of a square are the equilibrium points $(p,q)=(\pm 1,\pm 1)$.
Each slice has a small but finite thickness corresponding to the instanton width.}
\label{timeslices}
\end{figure}
The semi-classical prescription requires that one obtain all multi-instanton solutions of the classical EOMs which satisfy the BC's and add up the corresponding amplitudes.
	A typical multi-instanton configuration has been picturized in Fig.~\ref{timeslices}. 
	If the number of $P,Q,R$ instantons is $N=n+m+l$ then, for sufficiently well separated instantons, the classical action is additive,
	$S_{0}=nS_0^P+mS_0^Q+lS_0^R$ and the number of possible combinations is, 
	\begin{equation}
		\frac{N!}{n!\,m!\,l!}.
	\end{equation}
	This fully takes care of the purely classical part of the action - no functional integral had to be performed here.
	The quadratic fluctuations need more thought. Let $\mathbf{U}=\{U_1,\cdots U_N\}$ 
	be the disjoint, time ordered collection of $N$ time intervals, each interval being roughly one instanton wide, i.e. where $p^2,q^2$ differ substantially from one.
	Then the complement $\mathbf{\bar{U}}$ is the union of those intervals where $p^2\approx q^2\approx 1$, i.e. where the time evolution occurs via the SHO Hamiltonian for the $p,q$ DOFs. Obviously $\mathbf{U} \cup \mathbf{\bar{U}}=[-T/2,T/2] $. 
	\subsection{Graph-Theoretic Derivation}
\begin{figure}[H]
		\centering
		\includegraphics[scale=.35]{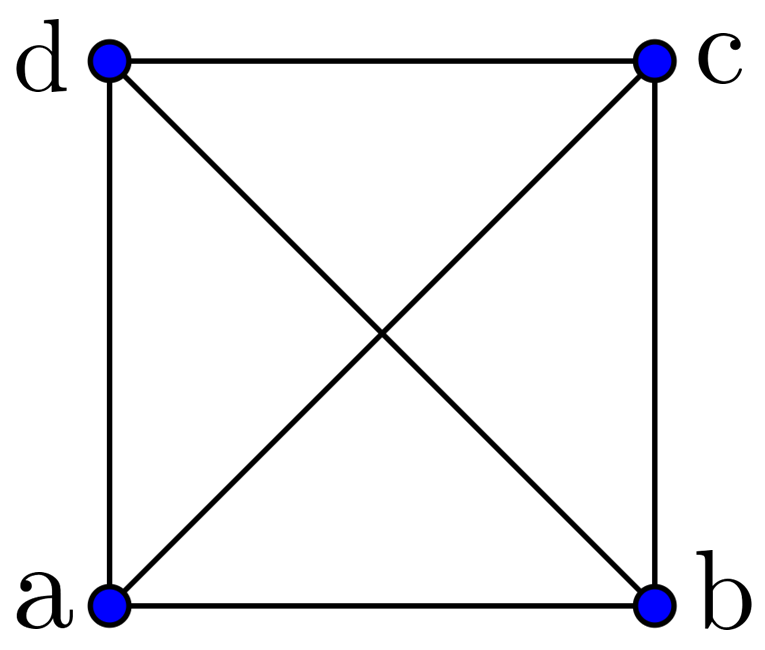}
		\caption{Horizontal, vertical, and diagonal transitions correspond to P,Q,R instantons. In graph theoretic language this is known as a $K_4$ graph.
			Our goal is to derive the transition amplitude between vertices.}
		\label{K4}
	\end{figure}

	The dilute gas approximation assumes that a vast distance separates one energy packet from the next and that, correspondingly, the fluctuations around one instanton cannot have any effect on the other.
	The action Eq.~\ref{Pot} is symmetric under $p\rightleftharpoons -p$ and $q \rightleftharpoons -q$, and so the transition amplitude from any one initial vertex in the $p-q$ plane to any other vertex (within the same time slice) is independent of the particular starting vertex.
	This means we can limit our attention to any one chosen vertex and consider horizontal, vertical, and diagonal transitions to the other three vertices (Fig.~\ref{K4}).
	Let $P_i,P_f$ be column/row vectors corresponding to the initial and final states respectively and $\mathbb{K}$ be the matrix below, 
	\begin{equation}
		\mathbb{K}= \begin{bmatrix}
			0 & K_Q & K_R & K_P \\
			K_Q & 0 & K_P & K_R \\
			K_R & K_P & 0 & K_Q \\
			K_P & K_R & K_Q & 0
		\end{bmatrix}.
		\label{K}
	\end{equation}
	In the language of graph theory this is the weighted adjacency matrix of the graph $K_4$ (fully connected with four vertices and six edges).
	The element $K_{ij}$ is the term contributed by the single instanton that connects minimum $i$ to minimum $j$.
	As an example choose $P_i=(1\;0\;0\;0\;)^T$, $P_f=(0\;1\;0\;0\;)^T$ for which $P_f^T \mathbb{K}P_i=K_Q$.
	The complete amplitude matrix can be constructed from the basic quantum mechanical rule of multiplying together amplitudes along a particular path.
	So, at the next time instant, i.e. when the second instanton ``fires'', the amplitude in Eq.~\ref{K} will be multiplied by $\mathbb{K}$ until the $N$'th one is reached. In fact $\mathbb{K}$ is just $E_0 \mathbb{I}-\mathbb{H}$, where $E_0$ is the energy in the absence of tunneling and $\mathbb{H}$ is the Hamiltonian in the $4\times 4$ truncated space. Each $\mathbb{K}$ matrix occurs sequentially, i.e. the instantons are time ordered by the identity, 
	\begin{equation}
		\int_{-\frac{T}{2}}^{\frac{T}{2}}\mathbb{K}dt_1\int_{t_1}^{\frac{T}{2}}\mathbb{K}dt_2 \cdots \int_{t_{N-1}}^{\frac{T}{2}}\mathbb{K}dt_N 
		=\frac{(\mathbb{K}T)^N}{N!}.
	\end{equation}
	When the above is summed over all $N$ we get, of course, $ e^{\mathbb{K}T}.$ Since $\mathbb{K}$ is a real symmetric matrix with a non-vanishing determinant it can be diagonalized and the amplitude matrix becomes $\mathbb{A}= e^{\mathbb{K}T}=S^{-1}e^{\Lambda T} S$ where 
	$S$ diagonalizes $\mathbb{K}$, i.e. $S^{-1}\mathbb{K}S=\Lambda$ where, 
	\begin{eqnarray}
		\Lambda= \begin{bmatrix}
			\lambda_S & 0 & 0 & 0 \\0 & \lambda_Q & 0 & 0 
			\\
			0 & 0 & \lambda_P & 0 \\0 & 0 & 0 &\lambda_R \\
		\end{bmatrix},\quad
		S=\frac{1}{2}\begin{bmatrix}
			1 & 1 & 1 & 1 \\
			-1 & 1 & -1 & 1 \\
			-1 & -1 & 1 & 1 \\
			1 & -1 & -1 & 1 \\
		\end{bmatrix}.
	\end{eqnarray}
	The eigenvalues of $\mathbb{K}$ are, 
	\begin{eqnarray}
		\lambda_S&=&K_P+K_Q+K_R,\quad
		\lambda_Q=-K_P+K_Q-K_R,
		\\
		\lambda_P&=&K_P-K_Q-K_R,\quad
		\lambda_R=-K_P-K_Q+K_R.
	\end{eqnarray}
	That $\sum \lambda_i=0$ follows from Tr $\mathbb{K}=0$.
	If the vertices of the square, i.e. the minima of the potential, are labeled $a,b,c,d$ (Fig.~\ref{K4}) then the tunneling amplitudes between them are, 
	
	\begin{equation}
		\begin{bmatrix}
			\mathcal{A}_{aa}\\ \mathcal{A}_{ab}\\
			\mathcal{A}_{ac}\\ \mathcal{A}_{ad}
		\end{bmatrix}
		= C \begin{bmatrix}
			c_P c_Q c_R + s_P s_Q s_R \\
			c_P c_R s_Q + c_Q s_P s_R \\
			c_R s_P s_Q + c_P c_Q s_R \\
			c_Q c_R s_P + c_P s_Q s_r 
		\end{bmatrix}
		= \frac{C}{4} 
		\begin{bmatrix} 
			1 & 1 & 1 & 1 \\ 
			1 & -1 & 1 & -1 \\ 
			1 & -1 & -1 & 1 \\ 
			1 & 1 & -1 & -1 
		\end{bmatrix} 
		\begin{bmatrix} 
			e^{\lambda_S T} \\ 
			e^{\lambda_P T} \\ 
			e^{\lambda_Q T} \\ 
			e^{\lambda_R T} 
		\end{bmatrix}.
		\label{finalamp}
	\end{equation}
	In the above we have defined $c_i \equiv \cosh(K_i T)$ and $s_i \equiv \sinh(K_i T)$ for $i \in \{P, Q, R\}$.
	By symmetry $\mathcal{A}_{ad} = \mathcal{A}_{bc},\; \mathcal{A}_{bd} = \mathcal{A}_{ac},\; \mathcal{A}_{cd} = \mathcal{A}_{ab}.$
	The above result Eq.~\ref{finalamp} generalizes the tunneling amplitude in the dilute instanton model for a single DOF to three DOFs.
	In order to get a better feel for this result consider the following special cases: 
	\begin{enumerate}
		\item \textbf{Horizontal/Vertical Instanton:} The $Q$ instanton (horizontal) corresponds to $K_P=K_R=0$.
		The only non-zero amplitudes are, 
		\begin{eqnarray}
			\begin{bmatrix}
				\mathcal{A}_{aa}\\ \mathcal{A}_{ab}
			\end{bmatrix}
			=C\left[
			\begin{array}{c}
				\cosh K_Q T \\
				\sinh K_Q T 
			\end{array}
			\right].
		\end{eqnarray}
		The vertical case is identical, with the $Q$ label exchanged for $P$.
		\item \textbf{Diagonal Instanton:} For this case $K_P=K_Q=0$ and the only non-zero amplitudes are, 
		\begin{eqnarray}
			\begin{bmatrix}
				\mathcal{A}_{aa}\\ \mathcal{A}_{ac}
			\end{bmatrix}
			=C\left[
			\begin{array}{c}
				\cosh K_R T \\
				\sinh K_R T 
			\end{array}
			\right].
		\end{eqnarray}
		\item \textbf{Equal Edge Instantons:} For $p,q$ fields with identical parameters $K_P=K_Q=K$ and $K_R=0$, all amplitudes are non-vanishing: 
		\begin{eqnarray}
			\begin{bmatrix}
				\mathcal{A}_{aa}\\ \mathcal{A}_{ab}\\
				\mathcal{A}_{ac}\\ \mathcal{A}_{ad}
			\end{bmatrix}
			=C\left[
			\begin{array}{c}
				\cosh^2 K T  \\
				\frac{1}{2}\sinh 2K T\\
				\sinh^2 K T\\
				\frac{1}{2}\sinh 2K T
			\end{array} \right].
		\end{eqnarray}
	\end{enumerate}
	
	\subsection{Gap Energies and Tunneling}
	If the barrier around each of the four wells was infinitely high, a particle would forever remain confined within that well and there would be a four-fold degeneracy.
	In the semi-classical limit, this degeneracy is lifted by tunneling events.
	Since our system has $Z_2 \times Z_2$ symmetry with four degenerate classical vacua $|a\rangle, |b\rangle, |c\rangle,$ and $|d\rangle$, the transition amplitudes calculated via the Euclidean path integral over a large time $T$ contain the complete information regarding the energy spectrum.
	The Euclidean evolution operator $e^{-HT}$ acting on states projects out all higher excited states.
	As $T \to \infty$, terms scaling as $e^{-E_n T}$ for $n \ge 1$ vanish relative to the ground state.
	This rigorously isolates the finite $4 \times 4$ subspace of the nearly degenerate ground states localized in the minima.
	Eigenstates of the Hamiltonian are the following parity-adapted combinations: 
	\begin{eqnarray}
		|\psi_{S}\rangle &=& \frac{1}{2} \bigg[ |a\rangle + |b\rangle + |c\rangle + |d\rangle \bigg],\quad
		|\psi_{Q}\rangle = \frac{1}{2} \bigg[ |a\rangle - |b\rangle - |c\rangle + |d\rangle \bigg],\nonumber \\
		|\psi_{P}\rangle &=& \frac{1}{2} \bigg[ |a\rangle + |b\rangle - |c\rangle - |d\rangle \bigg],\quad
		|\psi_{R}\rangle = \frac{1}{2} \bigg[ |a\rangle - |b\rangle + |c\rangle - |d\rangle \bigg].
		\label{psis}
	\end{eqnarray}
	The corresponding energies are $E_S=-\lambda_S,E_Q=-\lambda_Q,E_P=-\lambda_P,E_R=-\lambda_R$. Since we have specialized to the identical parameters case $K_P=K_Q=K$, there are only two energy gaps as measured from the lowest (symmetric) state, 
	\begin{equation}		
		\Delta E_P = 
		\Delta E_Q=E_P - E_S = 2(K + K_R), \quad 
		\Delta E_R = E_R - E_S = 4K.
		\label{split}
	\end{equation}
	\subsection{Quantum Versus Semi-Classical} 
\begin{figure}[H]
		\center
		\hspace*{-0.5cm} 
		\includegraphics
		[scale=.46]{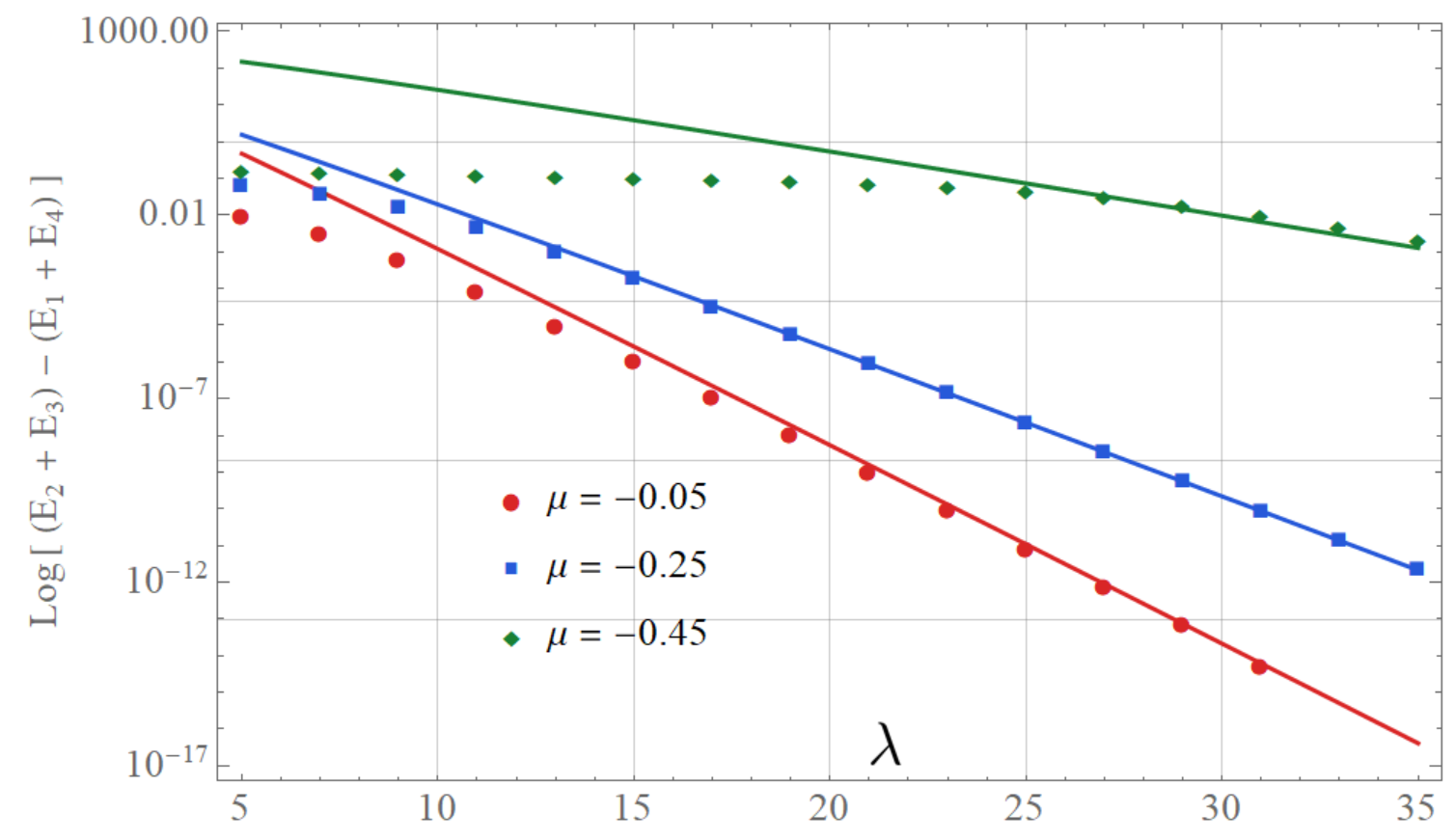}
		\caption{Semi-classical instanton results compared with high-precision numerical diagonalization for the ground-state energy splittings as a function of the scaling parameter $\lambda$ for the case of attractive coupling. The particular energy combination has been chosen to isolate the contribution for the $R$ (diagonal) instanton. $R$ exists only for $-\frac{1}{2}<\mu<0$.}
		\label{negmu}
	\end{figure}

	The above theoretical splittings can be compared against the four energies calculated with high precision for negative $\mu$ by solving the Schr\"{o}dinger partial differential equation on a grid: $E_1,E_2=E_3,E_4$.
	The semi-classical and quantum splittings should converge as the scaling parameter $\lambda$ gets larger.
	In Fig.~\ref{negmu} one indeed sees that this is true, affirming the correctness of our approximations. We have chosen to look at the combination $E_2+E_3-E_1-E_4$ because, from Eq.~\ref{split}, this isolates the diagonal instanton's contribution $K_R$. We see, as expected, that the numerical and theory results converge for large $\lambda$, and also that a much larger $\lambda$ is needed as $\mu\rightarrow-\frac{1}{2}.$ This dovetails exactly with the discussion preceding Eq.~\ref{lbig} on the unexpectedly large magnitude of the quadratic fluctuations.
	
\begin{figure}[H]	
		\center
		\hspace*{-0.5cm}	\includegraphics
		[scale=.47]{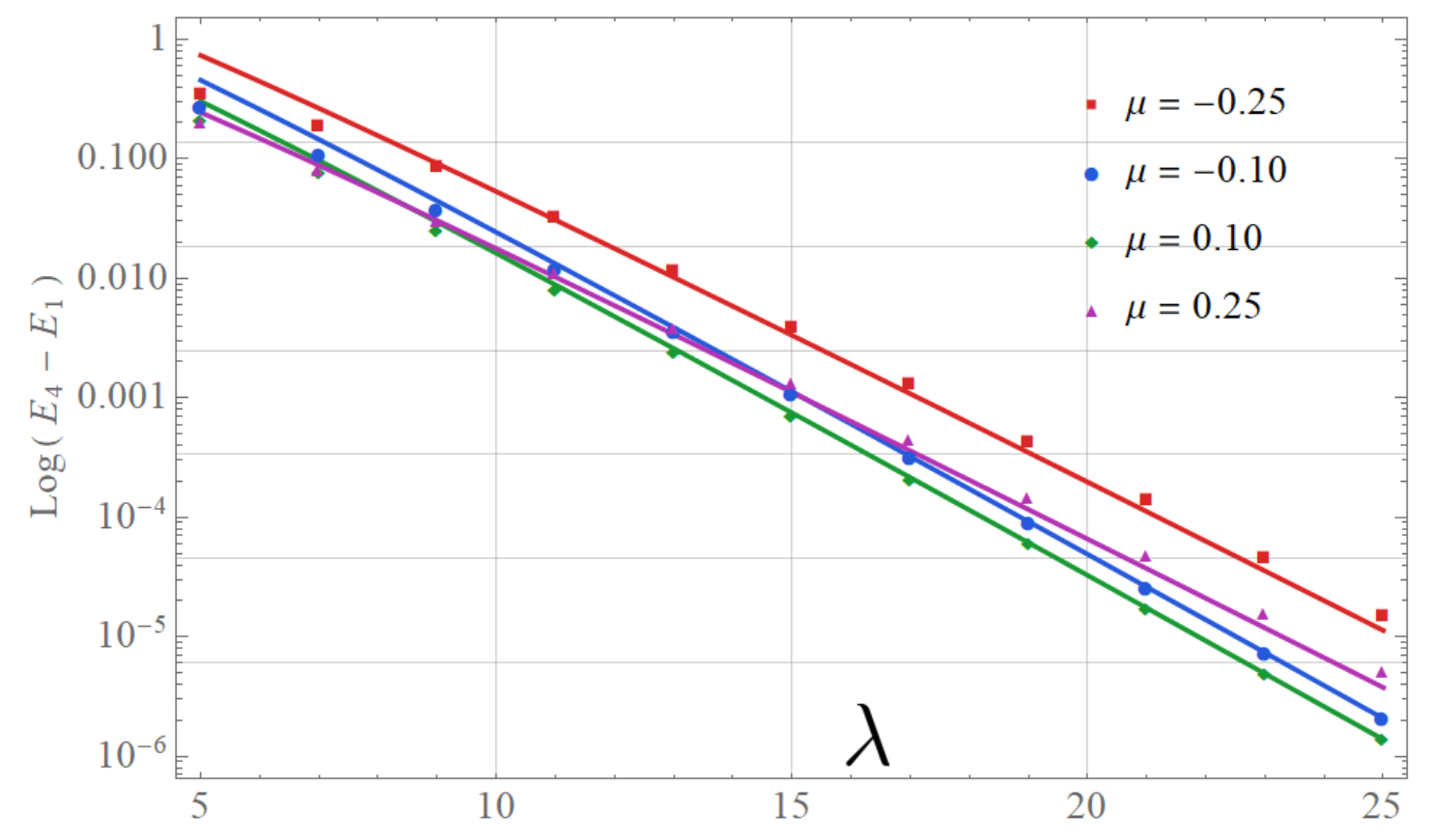}
		\caption{Comparison of high-precision numerical diagonalization with splittings computed perturbatively, $|\mu|\ll 1$. While the comparison in the large $\lambda$ limit is very good for small $\mu$, for larger $\mu$ the agreement expectedly becomes less perfect.} 
		\label{smallmu}
	\end{figure}

	Fig.~\ref{smallmu} compares the $E_R-E_S$ semi-classical and $E_4-E_1$ quantum splittings. This isolates the boundary hugging P-instanton, whose EOMs we were able to solve to $O(\mu^2)$ and fluctuation determinant to $O(\mu)$. As expected, while the comparison in the large $\lambda$ limit is very good for small $\mu$, for larger $\mu$ the agreement becomes less perfect. The primed and transverse determinant (see Eq.~\ref{d2}) show gentle behavior within the range explored unlike the violent behavior of $\chi_T^R$ at the boundary.
	
\begin{figure}[H]	
		\center	\hspace*{-0.5cm}	\includegraphics
		[scale=.5]{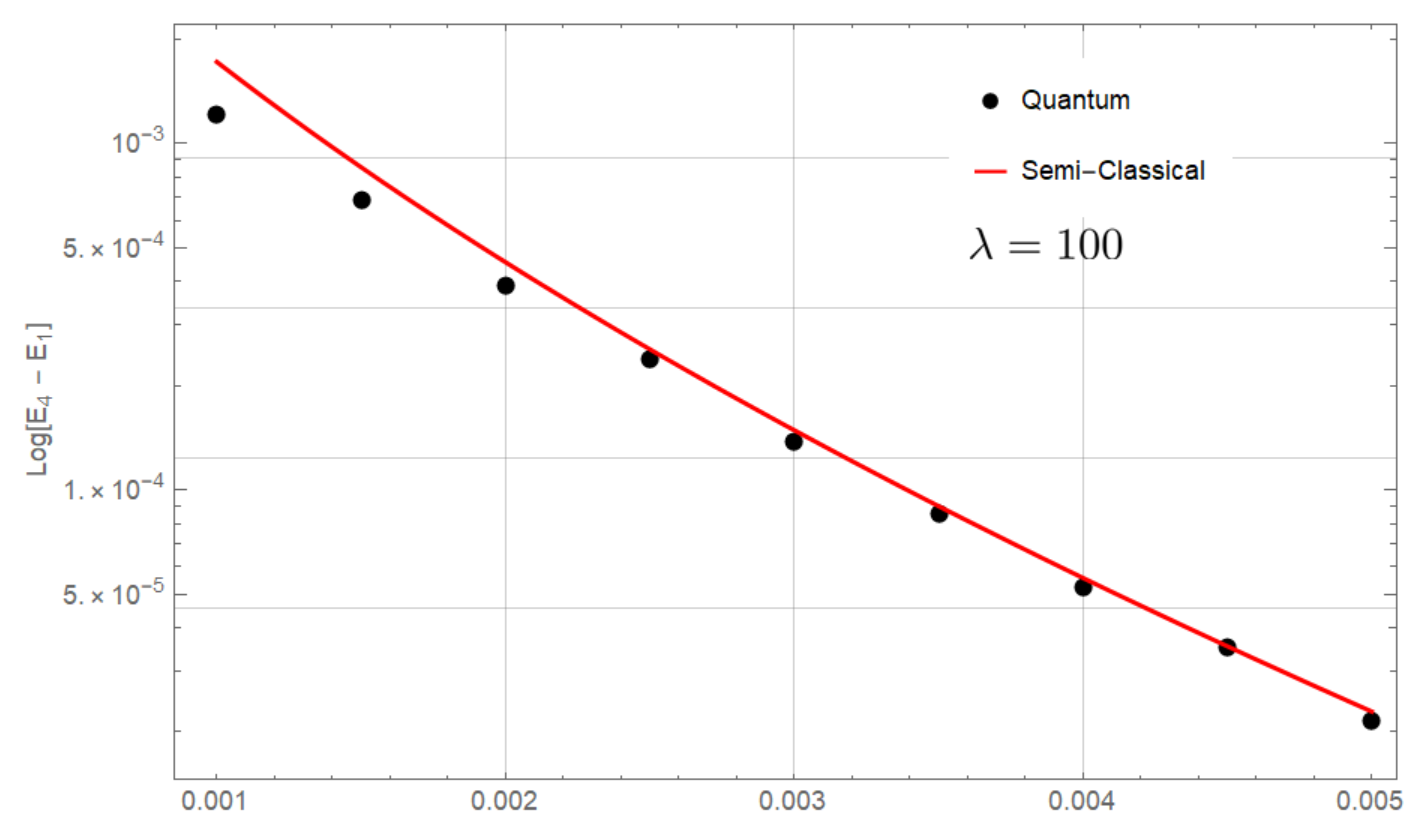}
		\caption{Quantum versus semi-classical results for fixed $\lambda$ and decreasing $\epsilon$. Although the mathematical formalism becomes more exact as $\epsilon=\frac{1}{2}-
			\mu\rightarrow 0$, the dilute gas approximation becomes progressively worse as instantons overlap.} 
		\label{mur}
	\end{figure}

	Finally, we turn to the extreme region $\mu\rightarrow \frac{1}{2}-\epsilon$. For $\mu>0$ we saw that there is no $R$ instanton and, on the other hand, closed form solutions for $P$ exist only for $\mu\ll 1$. This is why we explored solutions for $P$ near the right edge in Sections 5,6. Fig.~\ref{mur} compares quantum and semi-classical values up for fixed $\lambda$ and small values of $\epsilon$. Because we cannot increase $\lambda$ indefinitely for computational reasons, DIGA ultimately breaks down as $\epsilon\rightarrow 0$. 

	\subsection{Rabi Oscillations}
	
	To analyze the real-time dynamics of the quartic well, we perform a Wick rotation $\tau \to it$.
	The overall normalization constant $C$ drops out when calculating normalized probabilities, as the zero-point energy transforms into an unobservable global phase $e^{-iE_0t}$.
	Suppose that at $t=0$ the system is prepared entirely in the $a$ well, i.e. in $\ket{a}=\ket{-1, -1}$.	Thereafter the state evolves unitarily as $\ket{\psi(t)} = e^{-iHt}\ket{a}$. We can factor out the common global phase $e^{-i E_0 t}$, which vanishes when calculating probabilities. Thus the probability of remaining in well $a$ is,  	\begin{equation}
		P_a(t) = \frac{1}{4} \left[1+ \cos^2 2Kt +  2\cos2Kt\cos2K_Rt \right].
	\end{equation}
	To find the characteristic tunneling lifetime out of the initial minimum, we examine the short-time depletion rate: 
	\begin{equation}
		P_a(t)= 1 - (2 K^2 + K_R^2)t^2 + \mathcal{O}(t^4).
	\end{equation}
	Consequently, the characteristic lifetime $\tau$ is: 
	\begin{equation}
		\tau = \frac{\pi}{2 \sqrt{(2 K^2 + K_R^2)}}.
	\end{equation}
	The probability of transitioning from $a$ to adjacent minima $b$ or $d$ is, 
	\begin{equation}
		P_b(t) = P_d(t) = \frac{1}{4}\sin^2 2Kt.
	\end{equation}
	Finally, the probability of transitioning from $a$ to the diagonal minimum $c$ is, 
	\begin{equation}
		P_c(t) = \frac{1}{4} \left[ 1+\cos^2 2Kt - 2\cos 2Kt\cos2K_Rt\right].
	\end{equation}
	The dynamics reveal that tunneling to adjacent wells ($b, d$) is governed exclusively by the parameter $K$, while the probabilities of remaining in $a$ or tunneling diagonally to $c$ exhibit a beat frequency driven by the interplay between the orthogonal instanton paths $K$ and $K_R$.
	
	\section{Comparison with discrete approaches}
	
	Evaluation of the path integral via discretization is inevitable for practical reasons because analytical approaches become unworkable. The Ring Polymer Instanton (RPI) method \cite{Richardson1, Richardson2011, Lawrence2023, Richardson3, Mano, Kast, Kry,Erak, Benderskii} mentioned in the introduction provides a highly practical, computationally robust way to calculate tunneling in complex, multi-dimensional molecular systems. But, although it excels as a numerical algorithm, the present work has shown the importance of having an analytical formulation where possible. 
	
	First, time translation symmetry fully exists only in the continuum situation. Replacing differential equations with discrete difference equations explicitly breaks the fundamental symmetries of the system and turns exact translational zero-modes into pseudo-zero-modes \cite{Bender1, Munster}. Rigorous analytical control over discretization errors is lost \cite{Teper}.
	In RPI time translation is reduced to a discrete cyclic permutation group ($Z_N$). While this approximates the zero mode, it requires careful projection and finite-difference scaling. On the other hand, the canonical approach, time-translation symmetry yields an exact zero mode that is handled rigorously via the Faddeev-Popov procedure \cite{FP1,FP2, Zinn}, allowing for extraction of exact temporal volume factors.
	
	Second, the availability of an analytical solution yields insight into the tunneling amplitude and energy splittings. Because the action $S_0$ and the fluctuation determinant $\det \mathcal{M}$ are evaluated analytically, one can explicitly see how the tunneling splitting scales with physical parameters like the mass $m$, the barrier height $V_0$, or coupling constants like $\mu$. On the other hand, a numerical black box provides a specific number for a specific set of input parameters. It cannot provide a deep understanding of the functional dependence directly.
	
	Third, we recall that the instanton concept was invented to deal with certain problems of quantum field theory. Discretization is not an option here; extending the concept of discrete beads to higher-dimensional fields is computationally and conceptually unwieldy. The continuum approach alone was used to investigate topological defects in 4D spacetime such as BPST instantons and instanton-induced violations in QCD (the famous U(1) problem). Further, if a system possesses internal continuous spatial symmetries (e.g., $O(N)$ symmetric potentials), discretizing the path can introduce lattice artifacts that break these symmetries, complicating the rigorous extraction of the associated Goldstone modes. The continuous differential geometry used in the traditional approach is strictly necessary to evaluate the topological charges (Pontryagin indices) of these fields.

	\section{Summary and Conclusions}
	
	Extending the canonical single-field instanton framework to multi-degree-of-freedom spaces introduces severe mathematical and conceptual challenges. First, deriving the classical bounce trajectory requires solving coupled, non-linear Euler-Lagrange equations that govern the synchronous ``locking'' of interacting fields. Because such coupled systems generally evade closed-form solutions, standard treatments invariably fall back on discrete numerical approximations. 
	
	In this work, we have bypassed these approximations by exploiting the exact symmetries of a coupled quartic potential. By transforming the fluctuation operator into a comoving rotating frame, we explicitly accounted for the curvature-induced fictitious forces experienced by the interacting fields. This allowed us to cleanly isolate the longitudinal zero-mode and achieve an exact continuum evaluation of the transverse functional determinants. To ensure strict theoretical rigor in evaluating the Jacobian for the multidimensional zero-mode, we adapted the Faddeev-Popov procedure from gauge field theory to the quantum mechanical continuum. While this yields results consistent with standard 1D treatments, it provides reassurance that exact zero-mode extraction remains robust across coupled, multi-flavor instanton systems.
	
	A central result of our analytical approach is the derivation of transition amplitudes that do not artificially settle for a single path. By extending the dilute instanton gas model to encompass multiple topological sectors (two edge transitions and one diagonal transition), we mapped the infinite, time-sequenced instanton gas onto a weighted $K_4$ adjacency matrix. This graph-theoretic summation yielded exact closed-form expressions for the transition amplitudes, enabling the calculation of coherent Rabi-type oscillations between degenerate vacua. 
	
	The validity of our semiclassical derivations was affirmed by high-precision numerical diagonalization demonstrating good convergence. However, this convergence abruptly fails as the system approaches strong attractive coupling, $\mu \to -0.5$.
	
	\begin{equation}
		V(p,q)\bigg|_{\mu=-\frac{1}{2}}\sim(p^2-q^2)^2
	\end{equation}
	Classical particle motion must be either along the $p=\pm q$ line. Now there are no well-defined, isolated vacua. Instead the degenerate vacuum manifold is non-compact, consisting of intersecting continuous valleys. While there is no symmetry change in this limit, i.e. the collapsed potential Eq.~\ref{coll} retains the discrete $D_4$ symmetry of the full potential, 
	in this limit, the diagonal potential barrier vanishes and the four discrete degenerate minima dissolve.
	Crucially, this is reflected in our exact evaluation of the fluctuation determinant $\chi_T(\mu)$, providing a rigorous mathematical signature of this breakdown. The transverse prefactor develops an essential singularity, $\chi_T \sim \exp(4 \ln 2 / \sqrt{\epsilon})$. In normal tunneling calculations, the semiclassical rate is dominated by the exponential action $e^{-S_0}$, while the prefactor plays a subsidiary role. Here, the unphysical divergence of the prefactor explicitly signals that isolated, discrete quantum tunneling has ceased. Because discretized numerical pathfinders cannot natively resolve an essential singularity, this phenomenon establishes a hard theoretical boundary where computational instanton methods are structurally blind. Finally, note that this singularity is extracted from the exact closed-form evaluation of the diagonal sector (rather than a truncated perturbative expansion). Hence we can definitively establish that this breakdown is a fundamental feature of the continuum theory rather than an artifact of approximation.
	
	The lesson learned at the right edge, i.e. $\mu \to 0.5$, is interestingly different. The dilute instanton gas also loses validity here but the fluctuations remain modest sized and, in contrast to the situation on the left edge, the prefactor does not dominate over the exponential action. 
	
	While the immediate motivation for this exact model arose from the dynamics of composite molecular tunneling (Appendix B), the mathematical formalism derived here has broad relevance. Synchronous multi-field tunneling is a vital consideration in various physical regimes. For instance, in condensed matter systems, a two-dimensional free electron gas subjected to a spatially varying, perpendicular magnetic field forces electrons along snake-like classical paths. For spinless electrons, this is governed by a symmetric double-well potential \cite{Hoodbhoy}. Incorporating the electron's magnetic moment yields an effective action structurally isomorphic to the coupled multidimensional fields explored here. It is our hope that exact analytical frameworks like the one presented here will find further utility in modeling quantum transport, false vacuum decay, and the topology of suitably designed optical lattices.

	\appendix
	
	\section{Faddeev-Popov Procedure}	For a single DOF instanton textbooks give the standard method for extracting the Jacobian associated with the zero mode.
	On the other hand the Faddeev-Popov procedure \cite{FP1,FP2}, which was applied to the single DOF case by Zinn-Justin \cite{Zinn}, extends directly to the present case of multiple DOFs.
	To ensure rigor, we chose to go this way rather than assume the canonical prescription.
	The starting point is the identity,	 \begin{equation}		\frac{1}{\sqrt{2\pi\beta}}\int_{-\infty}^{\infty}d\lambda\;e^{-\frac{\lambda^2}{2\beta}}=1,\label{unity}
	\end{equation} 
	For the problem at hand, we make the following particular choice for $\lambda$:	 \begin{equation}
		\lambda(t^*)=\int dt\;\dot\Phi_c^T(t)A[\Phi(t+t^*)-\Phi_c(t)].\label{lam}
	\end{equation}
	Here $t^*$ is chosen arbitrarily with the intent of breaking the invariance of the action under time translations.
	The arbitrary parameter $\beta$ will eventually disappear from the final result for $\mathcal{A}$.
	The above expression is then inserted into the amplitude 
	\begin{eqnarray}
		\mathcal{A}&=&\mathcal{N}\int [d\Phi]\;  e^{-S[\Phi]}\times \frac{1}{\sqrt{2\pi\beta}}\int_{-\infty}^{\infty}
		dt^*\frac{d\lambda}{dt^*}\;e^{-\frac{\lambda^2}{2\beta}}\label{Ndef}.
	\end{eqnarray}
	Next, the integration variable is changed from $\Phi(t)$ to $R(t)=\Phi(t+t^*)$ and then back to $\Phi(t)$ making the integrand independent of $t^*$.
	After these manipulations the amplitude becomes, 
	\begin{eqnarray}
		\mathcal{A}&=&\mathcal{N}\int [d\Phi]
		\;\int \frac{dt^*dt}{\sqrt{2\pi\beta}}
		\dot\Phi_c^T(t)A\dot \Phi(t)\;e^{-S_\beta[\Phi]},\nonumber
		\\ \text{where, } S_{\beta}&\equiv&S+\frac{\lambda^2}{2\beta}\nonumber=S+\frac{1}{2\beta}
		\left[\int dt \dot\Phi_c^T(t)A[\Phi(t)-\Phi_c(t)]\right]^2.
		\label{AA}
	\end{eqnarray} 
	Since the second term above is positive definite, the effective action $S_\beta$ is obviously minimized, as was the original action $S$, at $\Phi=\Phi_c$.
	This leaves the EOMs unchanged and eliminates the linear term.
	The $t^*$ integral extends over the entire domain $[-T/2,T/2]$ and trivially gives $T$. Thus, 
	\begin{eqnarray}
		\mathcal{A}=
		\frac{\mathcal{N}T}{\sqrt{2\pi\beta}}
		\int [d\Phi]
		\;\int dt
		\dot\Phi_c^T(t)A\dot \Phi_c(t)
		e^{-S_\beta[\Phi]}.\;\;\;
	\end{eqnarray}
	At this point the $\Phi$ integration is replaced by integration over the set of basis coefficients $\alpha_n$ with $n=0,1,2\cdots$.
	\begin{eqnarray}
		\Omega(t)=\sum_{n=0}\alpha_n\Phi_n(t),\quad
		[d\Phi]
		=\frac{d\alpha_0}{\sqrt{2\pi}}[d\alpha]',\;\text{ where, }
		[d\alpha]'\equiv\prod_{n=1}^\infty\frac{d\alpha_n}{\sqrt{2\pi}}.
	\end{eqnarray}
	The boundary conditions at $\pm T/2$ ensure that, 
	\begin{equation}
		\int dt\;
		\dot\Phi_c^T(t)A\Phi_c(t)=0
	\end{equation}
	so that, after using Eq.~\ref{lam}, the parameter $\lambda(0)$ can be rewritten as $\sqrt{S_0}\alpha_0$.
	Performing the $\alpha_0$ integral then yields, 
	\begin{equation}
		\mathcal{A}=\mathcal{N}\frac{T}{\sqrt{2\pi}}e^{-S_0}\sqrt{S_0}\int [d\alpha]'\; e^{-S_2}
		\label{final},
	\end{equation}
	where the zero mode has been integrated out and the integration is now to be done over the quadratic level fluctuations.
	In the functional integration one may equally well integrate over fields in the lab fixed frame or the rotating frame.
	In the latter case the quadratic level part takes the form, 
	\begin{equation}
		\frac{1}{2}\int dt\;\mathcal{J} \tilde\Phi^T\mathcal{M}_{rot}
		\tilde\Phi.
	\end{equation} 
	
	We can easily check that the Jacobian $\mathcal{J}$ is unity: Let $\Phi(t)$ represent the fluctuation field in the fixed laboratory frame, and $\overline{\Phi}(t)$ represent the field in the rotating frame. They are related by a local, time-dependent rotation matrix $O(t) \in SO(N)$ with $\det(O(t)) = 1$:
	\begin{equation}
		\Phi(t) = O(t) \overline{\Phi}(t)
	\end{equation}
	The path integral measure transforms according to:
	\begin{equation}
		[d\Phi] = \mathcal{J} [d\overline{\Phi}]
	\end{equation}
	The functional Jacobian $\mathcal{J}$ for this continuous transformation is the infinite product of the determinants of the local transformation matrix evaluated at each instant in time. Hence,
	\begin{equation}
		\mathcal{J} = \prod_{t} \det(O(t))= \prod_{t} 1 = 1.
	\end{equation}
	Note, however, that the rotated matrices contain centrifugal and coriolis terms, and hence are different from the fixed frame matrices. 
	
	Denoting the eigenvalues of $A^{-1}\mathcal{M}_{rot}$ by $\lambda_n$, we can now complete the formal analysis: 
	\begin{eqnarray}
		S_2=\frac{1}{2}\sum_{n=0}\lambda_n\alpha_n^2.
	\end{eqnarray} and so the remaining integral is, 
	\begin{eqnarray}
		\int [d\alpha]'e^{-S_2}=
		\prod_{n\ne 0}
		\frac{1}{\sqrt{\lambda_n}}
		\equiv \frac{1}{\sqrt{\det' A^{-1}\mathcal{M}_{rot}}}.
	\end{eqnarray}
	(The $A^{-1}$ simply takes away the $a_p,a_q$ factors).
	For any classically defined path stable against transverse perturbations, the amplitude is, 
	\begin{eqnarray}
		\mathcal{A}=\frac{T}{\sqrt{2\pi}}e^{-S_0}\sqrt{S_0}	\frac{\mathcal{N}}{\sqrt{\det' A^{-1}
				\mathcal{M}_{rot}}}.\label{ZZ}
	\end{eqnarray}
	To complete the calculation we must determine $\mathcal{N}$.
	This will be done by taking the ratio with the fixed frame harmonic oscillator determinant - in this case that for two coupled oscillators.
	With reference to Fig. 1, let us consider the amplitude for a path that starts and returns from any one of the potential minima $(\pm 1,\pm 1)$ where the potential is nearly harmonic.
	In that case the semiclassical approximation to the amplitude gives, 
	\begin{equation}
		\mathcal{N}\int[dpdq]\;
		e^{-S_2}	=\frac{\mathcal{N}}	{\sqrt{\det A^{-1}\mathcal{M}_0}},
	\end{equation}
	Because it has only quadratic terms, the matrix $\det A^{-1}\mathcal{M}_0$ can be diagonalized and its determinant evaluated.
	The shifted frequencies are, 
	\begin{eqnarray}
		\omega^2_\pm &=&\frac{1}{2}(\omega_p^2+ \omega_q^2)\nonumber \pm \frac{1}{2}\sqrt{ (\omega_p^2-\omega_q^2)^2+16\mu\nu\omega_p^2\omega_q^2},
	\end{eqnarray}
	and the normalization constant is, 
	\begin{eqnarray}
		\mathcal{N}=
		\sqrt{\frac{m_p\omega_+}{\pi}}\sqrt{\frac{m_q\omega_-}{\pi}} e^{-\frac{1}{2}
			(\omega_+ + \omega_-)} \sqrt{\det A^{-1}	\mathcal{M}_0}
		\equiv C
		\sqrt{\det A^{-1}\mathcal{M}_0}.
	\end{eqnarray}
	For $\omega_p=\omega_q=\omega $ and $\nu=\mu,  
	\omega_\pm^2= \omega^2(1\pm 2\mu). \; C$ is defined in terms of the two particle harmonic oscillator wavefunction evaluated at the origin, 
	\begin{eqnarray}
		C=|\Psi_0(0,0)|^2 e^{-\frac{1}{2}
			(\omega_+ + \omega_-)T}.
		\label{CC}
	\end{eqnarray}
	Inserting the into Eq.~\ref{ZZ} we arrive at the final form of $\mathcal{A}$, which is the central result of this section, 
	\begin{equation}
		\mathcal{A}=CKT, \quad K=\sqrt{\frac{S_0}{2\pi}}e^{-S_0}	\sqrt{\frac{\det A^{-1}\mathcal{M}_0}{\det' A^{-1}\mathcal{M}_{rot}}}.\label{ZZZ}
	\end{equation}
	
	\section{Composite Tunneling}
	In atomic and nuclear physics the tunneling of a composite system has often been tackled using a
	coupled channel approach for scattering processes.
	This seeks to directly solve the time dependent Schr\"{o}dinger
	equation and is heavily computational. Though straightforward in principle, little theoretical insight can be gained. Some papers are referenced in the book by Razavy \cite{Razavy}.
	
Instead, let us consider a system where the tunneling object is finite-sized rather than point-like, i.e. a one-dimensional diatomic ``molecule'' made of two distinguishable point-like atoms joined by a perfectly rigid rod of length $L$. 	If such a molecule is placed entirely within one well, it will seek to tunnel into the other well much as a point particle would. 
Let us assume the two atoms experience 
identical potentials $U(z)$, each having the form of a symmetric double well, 
\begin{equation}
U(z)=\frac{m\omega^2}{16a^2}(z^2-a^2)^2.
\label{U(x)}
\end{equation}
For simplicity take the atomic masses as equal, $m_1=m_2=\frac{m}{2}$. The centre of mass $y=\frac{1}{2}(y_1+y_2)$ 
is equidistant from the two constituents and the relative distance $y=y_1-y_2
$ is fixed at $L$.
The (Euclidean) Lagrangian is, 
\begin{eqnarray}
\mathcal{L}=\frac{m}{2}\dot y^2+\bar U, \quad
\bar U= U\left(y+\frac{L}{2}
\right)+U\left(y-\frac{L}{2}\right)
\label{Ubar}
\end{eqnarray}
The potential $\bar U$ is symmetrical under $y\rightarrow -y$, i.e. the cm may be located equally within either well.
The key observation is that $\bar U$ may be written as a symmetric double well with minima at $y=\pm y_0$ with $y_0\ne a$.
\begin{eqnarray}	\bar U= \frac{m \omega ^2}{8 a^2} \left(y^2- y_0^2\right)^2+ C,\quad
y_0=\pm a f,\quad f=\sqrt{1-\frac{3}{4}\frac{L^2}{a^2}}. 
\end{eqnarray}
$C$ is a constant that may be discarded. In the limit $L \rightarrow 0$, $\bar U$ reduces to $2U$.
The classical EOM following from $\mathcal{L}$ is, 
\begin{equation}
\ddot y_c= \frac{\omega ^2}{2 a^2}(y_c^2-y_0^2) y_c ,  \label{EOM}
\end{equation}%
with a modified instanton solution that interpolates between the shifted vacua 
that are now located at $\pm y_0$.
Note that for $L\rightarrow 2a/\sqrt{3}$ the double minimum becomes a pure quartic with a single minimum at $y_0=0$, 
\begin{equation}
\bar U\rightarrow\frac{m \omega ^2 }{8 a^2}y^4+\frac{1}{18} m \omega ^2 a^2.
\end{equation}
There is no instanton solution in the above limit. 
The condition of perfect rigidity can be relaxed by adding a kinetic term for relative motion as well as an extra potential chosen to constrain $x$ near $L$, now to be thought of as the length parameter determining the average length of the vibrating molecule,  
\begin{equation}
\mathcal{L}=\frac{m}{2}\dot y^2+\frac{m}{8}\dot x^2 + U_T.
\end{equation}
In terms of the double well potential $U$ in Eq.~\ref{U(x)} the new potential $U_T$ is, 
\begin{equation}
U_T=U(y+\frac{x}{2})+U(y-\frac{x}{2}) +\frac{m\Omega^2}{32L^2}(x^2-L^2)^2
\end{equation}
As $\Omega\rightarrow \infty$ the molecule becomes increasingly rigid with equilibrium points close to $x=\pm L$.
To proceed further, we find the minima of $U_T$ and then re-express it with appropriately defined constants, 
\begin{eqnarray}
U_T=\frac{m\tilde\Omega ^2}{32 L^2}
(x^2-x_0^2)^2+ \frac{m\omega^2}{8 a^2}(y^2-y_0^2)^2
+\frac{3m\omega^2}{16 a^2}(y^2-y_0^2) (x^2-x_0^2)+C \label{pot}
\end{eqnarray}	
The potential minima are at $x=\pm x_0,\;y=\pm y_0$ where $x_0,y_0$ and the remaining constants are: 	\begin{eqnarray}	x_0^2&=&\frac{ 1-\frac{2\omega ^2}{\Omega ^2}} {1-\frac{ 2\omega ^2 L^2}{
\Omega ^2 a^2}}L^2,\quad
y_0^2=\frac{1-\frac{3 L^2}{4 a^2}-\frac{\omega ^2 L^2} {2 a^2 \Omega ^2}} {1-
\frac{2\omega^2 L^2}{a^2 \Omega ^2}}\;a^2\\
\tilde\Omega ^2 &=&\Omega^2\left(1+\frac{\omega^2L^2}{4 a^2 \Omega ^2}\right), \quad
C=\frac{1}{8} m \omega ^2 L^2\frac{1-\frac
{\omega ^2}{\Omega ^2} -\frac{L^2
}{2 a^2}}{1-\frac{2 L^2 \omega ^2}{a^2 \Omega ^2}}\label{pot2}
\end{eqnarray}
The constant $C=E_{min}$ is irrelevant to the dynamics but has been listed above for completeness.
Note that as $\Omega\rightarrow\infty$ the equilibrium positions shift towards the free values, $x_0\rightarrow L, y_0\rightarrow f a.$ 
Non-rigidity allows for molecular vibrations.
At the other extreme, one could imagine the atoms to be either totally free or very loosely bound together.
In that case, intuitively speaking, one atom may tunnel to the other side sooner than the other resulting in a $180^\circ$ flip of the molecule's orientation when eventually both atoms cross over.
One expects that the transition probabilities for spin flip and no flip will be equal if both atoms experience exactly the same potential.
However if there is some small difference then one would have an asymmetric double well with a single true vacuum and two different transition probabilities.
In Eq.~\ref{pot} we have precisely the form of quartic potential which we sought to explore in this work;
the potential $V(p,q)$ emerged in a very natural way. 
	
\section*{Conflict of Interest}
The authors have no conflicts to disclose.
\section*{Data Availability Statement}
The data that support the findings of this study are available from the corresponding author upon reasonable request.
	
\section*{AI Use Disclosure}
AI (Gemini, a large language model trained by Google) was used at several stages of this work: literature search, debugging Mathematica and LaTeX codes, proper formatting of figures, and rechecking formulae first derived by hand. We take full responsibility for all content generated or refined with its assistance.

\end{document}